\documentclass{article}

\usepackage{arxiv}

\usepackage{upquote}

\usepackage[utf8]{inputenc} 
\usepackage[T1]{fontenc}    
\usepackage{hyperref}       
\usepackage{url}            
\usepackage{booktabs}       
\usepackage{amsfonts}       
\usepackage{amssymb}
\usepackage{nicefrac}       
\usepackage{microtype}      
\usepackage{lipsum}
\usepackage{graphicx}
\usepackage{mdframed}
\usepackage{amsmath}
\usepackage{float}
\usepackage{xcolor}
\usepackage{natbib}
\usepackage{hyperref}

\usepackage{framed,color,verbatim}
\definecolor{shadecolor}{rgb}{.9, .9, .9}
\definecolor{cyan}{rgb}{.0, .99, .99}

\graphicspath{ {./images/} }

\usepackage{url}
\usepackage[pdftex]{epsfig}

\usepackage{color}
\newcommand{\remove}[1]{ }

   {\snugshade\verbatim}%
   {\endverbatim\endsnugshade}

\title{C3DIR: A Deep Learning 3-Dimensional Cloud Property Retrieval Scheme for Passive Satellite Imagers}

\author{
Charles H. White
\thanks{Corresponding author \\
{\bf CIRA:} Cooperative Institute for Research in the Atmosphere, Colorado State University, Fort Collins, CO. \\
{\bf ECE:}  Electrical and Computer Engineering, Colorado State University, Fort Collins, CO.
}
\\ 
    CIRA\\
    \texttt{charles.white@colostate.edu} \\
\And
    Yoo-Jeong Noh\\ 
    CIRA\\
    \texttt{yoo-jeong.noh@colostate.edu}
\And
    John M. Haynes\\
    CIRA\\
    \texttt{john.haynes@colostate.edu}
\And
    Imme Ebert-Uphoff\\
    CIRA, ECE \\
    \texttt{iebert@colostate.edu}
}

\begin{document}
\maketitle
\begin{abstract}

Operational cloud products from passive imaging satellite instruments often struggle to characterize cloud vertical structure due to limitations of visible, near-infrared, and infrared remote sensing. Recent efforts have shown that artificial intelligence (AI) and machine learning (ML) methods can improve depictions of cloud vertical structure. We develop the \textbf{Cloud 3-Dimensional Imager Retrieval (C3DIR)}, a deep learning model that estimates 3-D cloud properties for multiple passive satellite imagers trained to match retrievals from the Earth Cloud Aerosol and Radiation Explorer (EarthCARE) ACM-CAP product. This work is aimed towards moving AI/ML 3-D cloud algorithms closer towards operational use. C3DIR predicts the occurrence water content of ice, cloud liquid, and rain along the imager line-of-sight and uses a voxel-level collocation approach to account for the misaligned viewing geometries of passive imagers and active profiling instruments. This precise collocation methodology allows for constructing vertical profiles using voxels contained by multiple imager pixels to facilitate comparisons with active profiling instruments. 

We perform several evaluations of C3DIR focused on different aspects of its estimated 3-D volumes. Qualitative case studies show that C3DIR can accurately depict multiple distinct overlapping cloud layers, albeit with some smoothing. Quantitative evaluations illustrate that C3DIR overall excels at hydrometeor detection which intuitively tends to be a function of water content. However, detection of voxels classified as liquid cloud remains difficult due to the their small geometric thickness, finer horizontal scale, and frequent tendency to be obscured or embedded within ice clouds. In general, water content estimation is reasonably accurate, yielding the best results in ice clouds but uncertainties remain for liquid and rain water content. This is due to the limited sensitivity of passive visible, near-infrared, and infrared measurements to within-cloud structure and the difficulty of constraining these quantities in active sensor measurements through optically thick and precipitating clouds. Column-integrated water paths are in tighter agreement with EarthCARE. C3DIR similarly excels in estimating cloud-top height, cloud-base height, and cloudy profile detection. Comparisons with the algorithms underpinning current NOAA operational products highlight several areas where C3DIR may offer improvement. Overall, these results demonstrate the potential for C3DIR to provide flexible 3-D output depicting vertically resolved cloud structure which can offer broader utility for aviation applications, numerical weather modeling, and climate research.

\end{abstract}

\section{Introduction \& Background}
\label{intro}

Clouds are a fundamental component of weather and climate processes affecting atmospheric radiation, precipitation, and the broader energy and water cycles \citep{bony2015clouds,ipccar6}. The impacts of clouds due to the reflection, absorption, and emission of radiation depend strongly on both their macrophysical and microphysical properties as well as their geographic location \citep{stephens2005review, oreopoulos2017cvs}. Despite their importance, many properties of clouds are difficult to characterize due to their horizontal heterogeneity \citep{cahalan1994fractional, Zhang2011horizontal}, occurrence in vertically layered structures \citep{Barker1999geometry,barker2008cloudsat}, and the influence of interacting radiative and microphysical processes \citep{bony2015clouds,stephens2005review}.

Improving our ability to characterize properties of clouds, particularly their vertically resolved structure, is important for a number of operational applications. Cloud vertical structure is directly relevant to aviation applications where cloud vertical extent \citep{noh2022cbhaviation,noh2024cloud3d}, as well as the presence and amount of of supercooled liquid water, can significantly impact operations \citep{cao2018icing,bernstein2005icing,jeck2001icing}. Cloud detection and optical depth are also highly relevant for short-term solar energy forecasting \citep{miller2018solar, paletta2023solar} where cloud motions on short time scales have strong impacts on downwelling solar insolation at the surface. Derived motion winds have strong positive impacts on numerical model prediction and necessitate accurate height assignment of cloud features that are tracked \citep{hoffman2024dmw, velden2009dmv}. Cloud detection \citep{heidinger2012nbcm, frey2020mvcm, hunderbein2023mask} is an important upstream step needed for clear-sky imaging applications such as surface characterization and aerosol detection and quantification \citep{bulgin2018errors, martins2002errors}.

Passive imagers are a critical observation source due to the high temporal and spatial resolution needed to support near-real time applications \citep{mecikalski2007satellites}. However, cloud vertical profiling from these instruments remains under-constrained by visible (VIS), near-infrared (NIR), and infrared (IR) spectra alone \citep{heidinger2020cvsbook}. Additionally, issues arising from oblique viewing angles in certain regions complicate the performance, interpretation, and evaluation of these products \citep{maddux2010geometry}. As a result, many operational products from passive imagers focus on cloud-top, column-integrated, or single-layer depictions of cloudy scenes.

Cloud radars and lidars provide the most direct constraints on the vertical structure and microphysical properties of clouds. Spaceborne missions with these instruments such as CloudSat/CALIPSO \citep{stephens2002cloudsat, winker2010calipso} provide long-term, globally-distributed datasets serving as essential references for cloud products from passive imagers. These active observations and derived products have become essential for the training and evaluation of passive imager cloud products including cloud detection, cloud-top and base height, cloud-top phase, and others. The recently-launched Earth Cloud Aerosol and Radiation Explorer (EarthCARE) mission \citep{illingsworth2015ec, wehr2023ectech,eisinger2024earthcare,kubota2026earthcare} carries, among other instruments, an advanced cloud profiling radar and high spectral resolution lidar that improves upon this valuable record of observations. However, the key limitation of these instruments is that they severely lack the spatial and temporal coverage needed for many real-time applications that is provided by passive imaging sensors. 

Despite the difficulty in assessing cloud vertical structure from passive VIS, NIR, and IR spectra, there are several physical and statistical approaches targeted towards characterizing vertically-resolved cloud properties or detection of the multiple distinct layers. \citet{watts2011oca} use an optimal estimation framework and handle multi-layer scenes by identifying high cost retrievals and reprocessing them with a simplified two-layer model. The Algorithm Working Group Cloud Height Algorithm (ACHA; \citealp{heidinger2020achaatbd}), which underpins the NOAA operational cloud-top height (CTH) product, also has multi-layer capability using a similarly simplified representation of lower cloud layers. 

Artificial intelligence and machine learning (AI/ML) models have become an increasingly common approach for characterizing cloud vertical structure. This is in part due to their flexibility and ability to encode statistical associations between spectral and spatial variability that are not easily accounted for in physical models. \citet{haynes2022lowcloud} demonstrated that a relatively simple random forest or neural network model can significantly improve the detection of low cloud layers under upper-level clouds in the NOAA operational Cloud Cover Layers (CCL) product. Several convolutional neural network (CNN) or transformer-based approaches take into account spatial information to infer the presence of clouds at high vertical resolution. \citet{wang2023cloud3d} use a CNN to estimate vertical cloud masks provided by the CloudSat 2B-CLDCLASS-LIDAR product \citep{sassen2008cldclasslidar} showing improvements over relatively simple multi-layer perceptrons (MLP) models. \citet{bruning2024cloud3d} train a similar U-Net to estimate CloudSat radar reflectivity. \citet{girtsou2025cloud3d} used a small vision transformer to similarly estimate CloudSat radar reflectivity and show significant benefit in using self-supervised pretraining before fine-tuning on the relatively sparse active sensor profiles. However, the radar reflectivity alone is likely to miss upper-level thin clouds that are more easily detected through lidar-only or combined radar-lidar retrieval \citep{comstock2002cirrus,mason2024comparison}.

Other 3-D AI approaches use combined lidar and radar retrievals of cloud properties as labels, specifically the DARDAR cloud products (raDAR/liDAR;\citealp {delanoe2010dardar}). DARDAR is a variational method independent of the NASA products, synergistically combining observations from CloudSat, CALIPSO, and MODIS to retrieve properties of ice clouds. In particular, it allows for more seamless characterization of clouds between regions detected by both active instruments and regions where only one is sensitive. \citet{jeggle2025cloud3d} developed IceCloudNet, a U-Net that estimates water content and number concentration of ice clouds with DARDAR serving as the reference. The Chalmers Cloud Ice Climatology \citep{amell2024cloud3d} is a similarly designed U-Net approach based on single-channel 11 µm imagery from GridSat \citep{knapp2011gridsat} and estimates ice water content and ice water path from DARDAR. DARDAR is more sensitive to upper-level thin clouds than radar reflectivity due to its use of lidar observations, but it lacks depictions of liquid clouds making it difficult to develop a general purpose imager cloud product that relies solely on DARDAR.

\subsection{Proposed Approach}
Building upon some aspects of previous AI/ML-based efforts, we develop a CNN that we call the Cloud 3-Dimensional Imager Retrieval (C3DIR; pronounced "cedar"). C3DIR attempts to address several aspects of previous efforts to characterize 3-D cloud structure from passive imaging instruments. C3DIR leverages both spatial and spectral variability to estimate the presence and vertical extent of cloud cover, characterize multiple distinct layers, and quantify the vertical variation of cloud water content and phase. Our efforts here are focused on pushing AI-based 3-D cloud retrievals toward operational applications, where robustness across different imagers and viewing conditions is essential and where significant benefit over more traditional operational approaches must be demonstrated.  

Key elements of our approach include: 
\begin{itemize}
\item
  {\bf 3-D collocation scheme:} Previous efforts have largely relied on matching imager pixels representing slant paths through the atmosphere to vertical profiles nearly normal to the surface, implicitly simplifying the 3-D geometry problem. Issues associated with this simplification are often mitigated by removing profiles with exceedingly high viewing angles (typically less than 45 degrees; e.g., \citealp{bruning2024cloud3d,jeggle2025cloud3d,girtsou2025cloud3d}). This severely limits the geographic areas viewed by geostationary imagers, does not completely solve the misalignment of observations, and potentially limits generalization capacity outside this region. We present a more precise collocation methodology used in C3DIR that directly accounts for the frequently misaligned views of these instruments and allows us to include oblique viewing angles in our training and evaluation data. This methodology is relatively simple, and directly accounts for common failure modes associated with matching entire active sensor vertical profiles to a single imager pixel. Our approach instead matches select voxels in C3DIR's output space to closely-matched portions of the active sensor profile. 
  See Section \ref{sec:collocation} for more details. 
\item
  {\bf Single AI model for multiple sensors:} C3DIR uses observations from the newly-launched EarthCARE satellite as labeled training and evaluation data. The recently published ACM-CAP \citep{mason2023acmcap} product uses EarthCARE observations from several instruments to provide a more complete depiction of cloud properties in the vertical profile. Despite EarthCARE being operational for almost two years, collocations between EarthCARE and geostationary imagers are relatively sparse representing only a narrow curtain transecting otherwise unlabeled passive imagery. A separate problem is that passive imagers often have differing channel availability, spectral response functions, and spatial resolutions which necessitate treating each sensor individually. We incorporate a module within C3DIR that takes into account the individual channel selection and spectral response of each satellite at a very early stage of the model. This architecture allows us to train a single model for all passive imagers included in the training data. This also effectively increases the size of our training dataset and allows a given sensor to benefit from collocations between EarthCARE and an entirely different imaging instrument. See Section \ref{sec:model} for more details.
\end{itemize}

Additionally, we utilize information from both satellites and Numerical Weather Prediction (NWP) models as input (Section \ref{sec:model_inputs}).
We also use auxiliary labels in a multi-task learning approach to provide additional supervision to the model (Section \ref{sec:auxiliary_labels}) outside of the EarthCARE ground track.

We first provide a basic description of the instruments and data used in Section \ref{sec:data}. Section \ref{sec:collocation} describes the 3-D collocation methodology. Section \ref{sec:model} provides an overview of the C3DIR model architecture including training details. Section \ref{sec:results} presents an evaluation of C3DIR predictions from several perspectives including intercomparisons with NOAA operational products, voxel-, profile- and layer-based evaluations, and comparisons with ground-based remote sensing instrumentation.  Sections \ref{sec:discussion} and \ref{sec:conclusions} present a discussion of the results and final conclusions. 

\section{Data}
\label{sec:data}

\subsection{Model Inputs}
\label{sec:model_inputs}

C3DIR is built to accommodate several different imaging instruments as inputs to the model. This specific version of C3DIR is trained to use observations from multiple geostationary imagers including three Advanced Baseline Imagers (ABI; \citealp{schmit2017abi}) aboard the NOAA Geostationary Operational Environmental Satellites (GOES)-R series (GOES-16, GOES-18, and GOES-19); two Spinning Enhanced Visible and Infrared Imagers (SEVIRI; \citealp{schmetz2002seviri}) aboard EUMETSAT's Meteosat Second Generation satellites (Meteosat-9 and Meteosat-10); and one Advanced Himawari Imager (AHI; \citealp{bessho2016ahi}) aboard Himawari-9 operated by the Japan Meteorological Agency (JMA).

All available VIS, NIR and IR channels on each instrument are used with central wavelengths spanning 0.47--13.3~\(\mu\mathrm{m}\). Since the native spatial resolution varies between channels, each channel is coarsened to the lowest resolution available on each imager. For ABI, this requires a \(4 \times 4\) average pooling to the 0.5 km Band-2 and \(2 \times 2\) average pooling to the three 1 km channels to bring all sixteen channels to a common 2 km spatial resolution. For the purposes of this study we only include the 10-minute full-disk imagery from ABI and AHI and the 15-minute full-disk imagery from SEVIRI in the training dataset. Note that the high resolution visible (HRV) channel on SEVIRI is not included due to the lack of full-disk coverage. 

In addition to the observations from the imaging instruments we also use some limited information from the closest 6-hour forecast from the Global Forecast System (GFS; \citealp{ncep2015gfs025}) at 0.25 degree resolution valid for the same time as the imager observations. Specifically, we use temperature and relative humidity from 1000 mb to 100 mb at 100 mb intervals along with the pressure and temperature at the surface and tropopause. 6-hour forecasts are chosen instead of the analysis in order to accommodate near-real-time use in operations. These model forecasts are used as input to the model to help approximate contributions from the clear-sky atmosphere and surface to the top-of-atmosphere (TOA) radiances observed by the imagers.

\subsection{Primary Labels}

EarthCARE is a recently launched satellite mission consisting of several instrument packages including the Cloud Profiling Radar (CPR), the ATmospheric LIDAR (ATLID), the Multi Spectral Imager (MSI), and the Broad-Band Radiometer (BBR). Similar to CloudSat and CALIOP, EarthCARE flies in a polar orbit allowing for a globally distributed dataset of high-quality near-nadir vertical profiles of the atmosphere. Throughout this work, we rely primarily on the synergistic ACM-CAP product \citep{mason2023acmcap}, where ACM denotes the ATLID, CPR, and MSI instrument synergy, and CAP denotes the retrieved clouds, aerosols, and precipitation. The ACM-CAP product combines measurements from theses instruments to estimate several vertically-resolved characteristics of clouds and aerosols. The ACM-CAP product relies on the Cloud, Aerosol, and Precipitation from mulTiple Instruemnts using a VAriational TEchnique (CAPTIVATE; \citealp{mason2017captivate, mason2018captivate}) framework, which is a flexible optimal estimation method providing, among many other fields, vertically-resolved properties of ice, liquid, rain.

Of primary interest to this work are the derived ice, liquid, and rain water content fields (hereafter abbreviated as IWC, LWC and RWC) in the ACM-CAP product which provide the concentration of water at a vertical resolution of 100 m throughout the troposphere on the 1 km horizontal Joint Standard Grid (JSG) which is EarthCARE's Level-2 common reference grid system \citep{eisinger2024earthcare}. To match the output resolution of C3DIR, these fields are averaged to a vertical resolution of 0.25 km from -0.125 km to 19.875 km. After averaging, we derive an additional four fields based on the binary detection of any concentration greater than \(1 \times 10^{-5}~\mathrm{g}~\mathrm{m}^{-3}\) of water: an ice mask, liquid mask, rain mask, and a fourth hydrometeor mask indicating the presence of any species of water. These vertically-coarsened water contents and masks serve as the primary source of labeled data for training and evaluating C3DIR.

Note that ACM-CAP ice water content encompasses precipitating and non-precipitating ice particles, but cloud liquid and rain are treated as separate quantities. C3DIR follows this same convention and does not combine cloud liquid and rain due to the differing optical extinction properties of rain resulting from their differing particle size distributions. Because clouds are the primary focus of C3DIR's target applications, we consider the rain mask and rain water content to be of secondary importance. However, we include them here so that C3DIR produces a more complete accounting of all hydrometeors in the atmospheric column. 

\subsection{Auxiliary Labels for Multi-Task Learning}
\label{sec:auxiliary_labels}

One challenge in using vertical profiles from EarthCARE to train a CNN is the severe sparsity of the labels relative to the size of the output space of the model. For a \(64 \times 64\) image cut-out from a passive imager, we find that typically less than 2.2\% of voxels contain a collocated label from EarthCARE ACM-CAP product. Thus, while the labels from ACM-CAP contain extremely valuable information regarding cloud vertical structure, they provide no direct supervision outside the ground-track of the the satellite.

We add weak constraints outside the EarthCARE ground track by using a multi-task learning approach with auxiliary labels.  Namely, we train C3DIR to additionally estimate three fields derived from the algorithms underpinning the NOAA operational products for ABI. These products are (1) a binary cloud mask derived from the Enterprise Cloud Mask (ECM; \citealp{heidinger2020ecm}); (2) the CTH from ACHA \citep{heidinger2020achaatbd}; and (3) a binary ice and liquid classification derived from the cloud-top phase algorithm \citep{pavolonis2020phase}. We find that including these labels outside the EarthCARE ground-track slightly improves the generalization performance on the validation set. These three products were chosen specifically since they are properties of clouds that passive imagers are especially sensitive to and are available at all times of day. We run these algorithms for all imagers included in this work within the CLAVR-x processing system, which serves as the development testbed for several of the NOAA operational cloud products. 

As stated above, the presence and water content of ice, and liquid are of primary interest to this work, but we also found benefit in including the ice and liquid effective radius from the ACM-CAP product as outputs. We hypothesize that this is because the detection of clouds in passive imagery is most often related to a cloud's optical depth, which is a function of both water content and particle size. Overall, the inclusion of ice and liquid effective radius improved the generalization performance of C3DIR for water content and detection. However, until  ACM-CAP itself undergoes more rigorous evaluation, we choose to leave an evaluation of the particle size estimates from C3DIR to future work. We make no claims regarding C3DIR's performance in estimating particle size with respect to ACM-CAP and, along with C3DIR's estimates of the NOAA operational products, we consider these auxiliary outputs to be purely experimental. 

\subsection{Additional Evaluation Data}

In addition to the EarthCARE ACM-CAP products, we also use the Continuous Baseline Microphysical Retrieval (MICROBASE; \citealp{wang2011microbase}) product available from ground-based remote sensing measurements at the U.S. Department of Energy's Atmospheric Radiation Measurement (ARM) User Facility at the Southern Great Plains (SGP) site located in the south central United States. This algorithm combines several products and observations from a radar, lidar, ceilometer, as well as a microwave radiometer estimate of liquid water path. The MICROBASE product provides estimates of liquid and ice water content every 4 seconds at 30 m vertical resolution. These products are similarly coarsened in the vertical dimension to match the same resolution as the coarsened ACM-CAP product and the output of C3DIR. We run C3DIR on the continental United States (CONUS) sectors from GOES-16 ABI available every 5 minutes to perform comparisons with the vertical profiles from the SGP-ARM site. The coarsened MICROBASE estimates are not used during training and serve as a secondary point of comparison of C3DIR's output. The data we use from this product covers the period from January 1, 2024 to June 29, 2024. 

\section{Collocation Methodology}
\label{sec:collocation}

\begin{figure}[t]
    \centering
    \includegraphics[width=16cm]{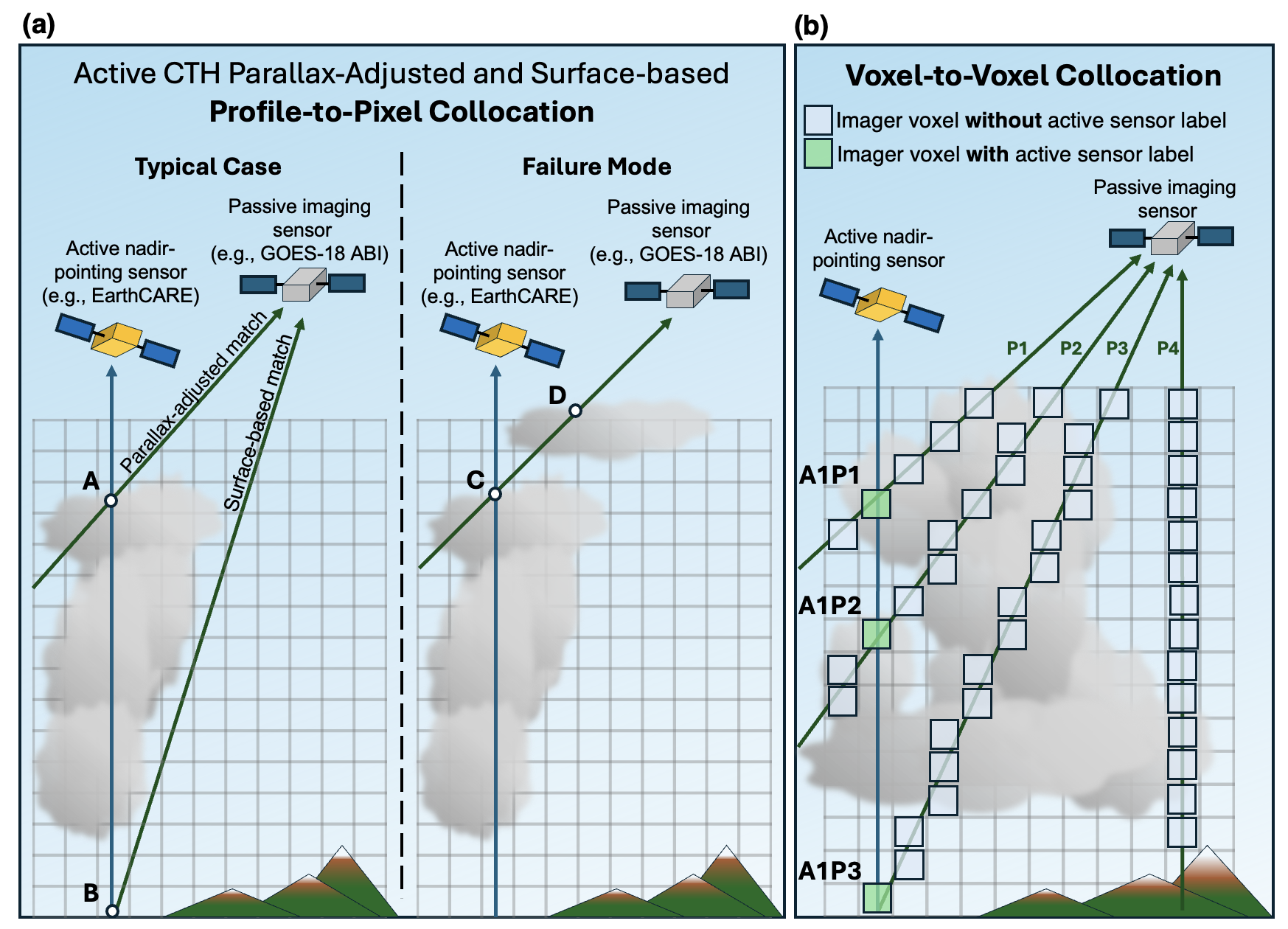}
    \caption{Conceptual diagram illustrating the geometry of the collocation methodology. (a; left-hand side) illustrates CTH parallax adjusted matching between imager pixels and active sensor profiles. A common failure mode of this approach is shown on the right-hand side of (a). Points A and C represent the CTH measured by the active sensor, Point B represents the intersection of the imager pixel and the active sensor profile at the surface, and Point D represents the CTH of a cloud observed by the imager but not seen by the active sensor. (b) Simplified illustration of proposed voxel-based collocation strategy. P1-P4 represent different imager pixels while A1P1-A1P3 represent individual voxels matched between the single active sensor profile and the corresponding imager pixel.}
    \label{fig1}
\end{figure}

The simplest approach for collocating passive imaging instruments and active sensor profiles is to collocate the two sets of observations based on their coordinates assigned by their intersection with the ellipsoid (Point B in Figure \ref{fig1}.a). However, since clouds are often elevated well above the surface this can lead to substantial error in attempting to evaluate cloud property algorithms due to misalignment of the views of each sensor. 

One common way of addressing this issue is to find the cloud-top identified in the active observations and select the imager pixel that intersects with the profile at that altitude (Point A in Figure \ref{fig1}.a). For the purposes of characterizing cloud-top properties, the CTH parallax-adjusted collocation procedure can account for some amount of the view misalignment. However, this only solves the problem to a first order and issues associated with view misalignment still remain. The right-hand side of Figure \ref{fig1}.a illustrates a common failure mode of this approach where a cloud is seen by the passive imager, but is not characterized in the active sensor observations. The key issue is that the vast majority of passive imaging sensors will not have collocations where the path the imager sees through the atmosphere is completely aligned with the active sensor. Even after parallax adjustment, both methods above (represented by Points A and B) yield sets of collocated observations that are only truly representative of each other at their intersection. As a result, we argue that tuning or evaluating passive imager cloud algorithms that characterize any vertical extent of the profile implicitly assumes some amount of local homogeneity in cloud properties in a horizontal area proportional to the viewing angle when matching to nadir-pointing active sensors in this manner.

While there are clear issues surrounding the approach depicted in Figure \ref{fig1}.a, it is still reasonable in many situations. Imaging instruments measuring TOA radiation see combined contributions from the atmosphere above the cloud, some depth within the cloud, and for optically thin clouds, the atmosphere and surface underneath them. One perspective in support of this methodology might be that for optically thick clouds, the weighting function describing contributions of TOA radiation in the visible to infrared range peak near cloud-top. We expect issues to arise mostly for optically thin clouds where the weighting function does not peak as sharply within the cloud, and for highly spatially heterogeneous cloud fields (e.g., right hand side of Figure \ref{fig1}.a). In many applications it may be reasonable to exclude such cases when evaluating physically-based algorithms. However, when tuning statistical models (such as large AI/ML systems) to match properties of active sensor profiles, excluding optically thin clouds, spatially variable scenes, or high viewing angle collocations may be detrimental to model generalization performance and severely limit the amount of training data that can be used.

\subsection{3-D Collocation Method}
To more directly solve the problem of misaligned views between these two sets of observations, we propose an alternative collocation methodology centered around collocating portions of the active vertical profile with the output space of a 3-dimensional cloud property retrieval: a voxel-to-voxel matching. This is in contrast to collocating based on the input space of the model which one might consider to be a profile-to-pixel matching.

Figure \ref{fig1}.b is a simplified illustration of this methodology. We take the passive imager pixels (P1-P4) and discretize them into voxels at fixed altitudes along the slant path the imager sees through the entire atmosphere down to the surface. Each voxel and its cloud properties are assigned a unique triplet of altitude, latitude, and longitude. For the purposes of tuning or evaluating the 3-D model, most voxels associated with a given pixel do not have a corresponding label from the active sensor and the vast majority of pixels do not have any labeled voxels at all (P4, for example). P1-P3 each have a single labeled voxel represented by A1P1-A1P3 in Figure \ref{fig1}.b. This set of three labeled voxels and their corresponding coordinates represent the training data used for C3DIR. In the example shown, C3DIR is tasked with estimating the presence and water content of clouds in all voxels (labeled or not), but is only penalized during training based on the collocated voxels, A1P1-A1P3. An important note is that in many cases there can be several imager pixels that contain voxels corresponding to different parts of a single active sensor profile. The number of such voxels depends on the passive sensor viewing angle, spatial resolution of the sensor, and the maximum altitude considered for collocation.

Figure \ref{fig1}.b illustrates a regular grid with the vertical dimension normal to the surface. This is primarily to ease visualization and in practice there is no need to create the voxels along this grid. When designing the dataset to train C3DIR, we calculate the latitude and longitude at fixed predefined altitude intervals along the path the imager sees down to the surface without resampling to a regular latitude/longitude grid. Therefore, the vertical dimension is slanted from the intersection at the surface up to the satellite following this path. The creation of a 3-D volume with the vertical dimension normal to the surface is relatively simple and can be done with a nearest-neighbor sampling at each altitude from multiple imager pixels as a post-processing step. Another difference between our approach and Figure \ref{fig1}.b is that we ensure that every vertical element of the active profile is matched with a corresponding passive imager voxel provided that its horizontal separation is less than 5 km.

Our overall expectation is that using this more precise collocation methodology will allow us to make better use of high viewing angle collocations and reduce the amount of model error due to misaligned views between nadir-pointing active sensors and passive scanning imagers. Note that we expect the voxel-to-voxel collocation scheme to be appropriate only when the output space of the model of interest is a vertical profile similar to the active sensor, and not when the output is a single value describing some aspect of the vertical profile e.g., CTH, cloud-base height (CBH), or water path.

\begin{figure}[t]
    \centering
    \includegraphics[width=16cm]{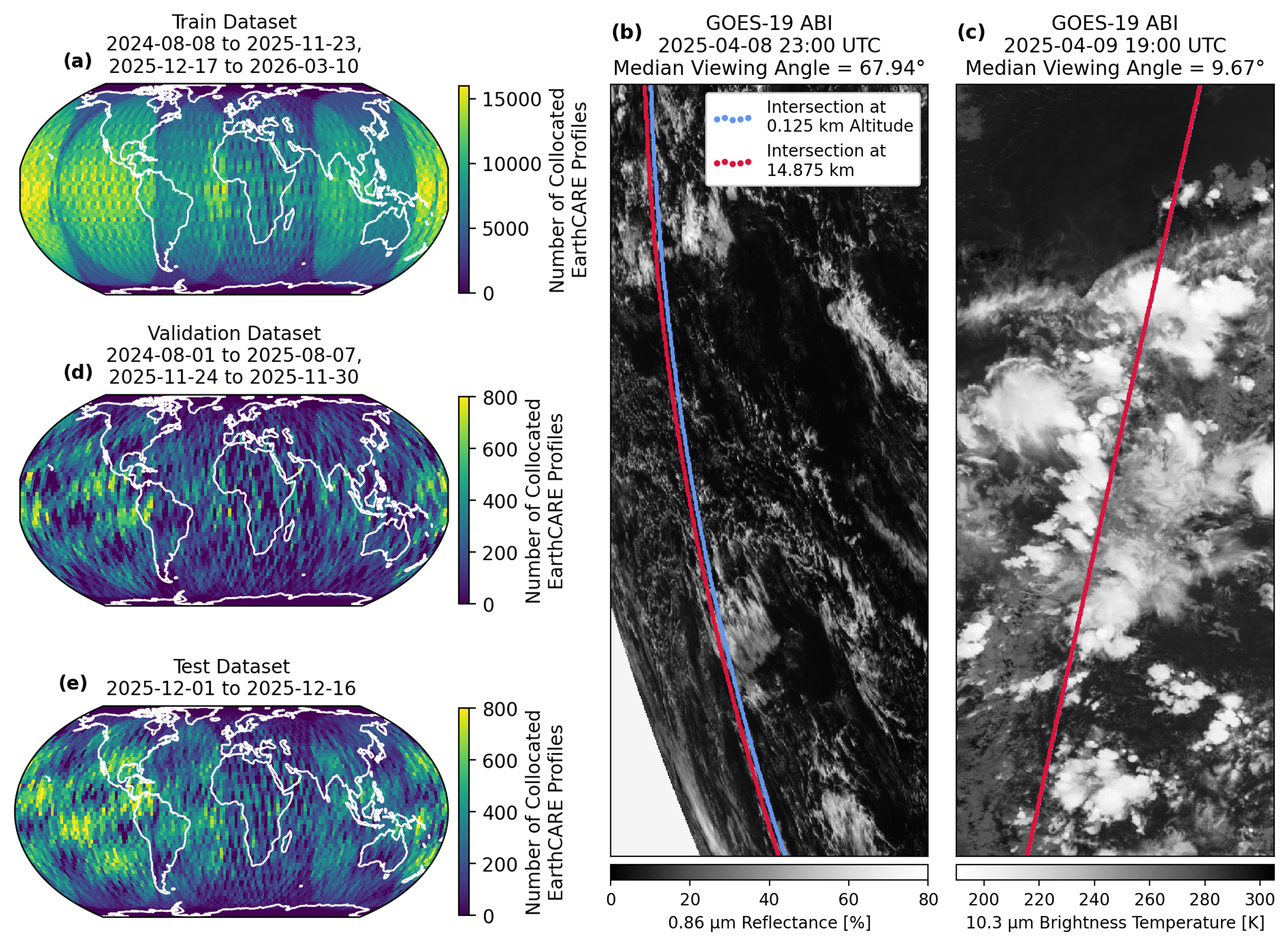}
    \caption{Global distribution of collocations and examples. (a), (d), and (e) show the distribution of dates of collocations included in the training, validation and testing datasets. (b) and (c) show the selected pixels from GOES-19 that are matched with an EarthCARE cross section at 0.125 km and at 14.875 km altitude. (b) shows an example for high viewing angles (roughly 68 degrees) and (c) shows this for much lower viewing angles (roughly 10 degrees).}
    \label{fig2}
\end{figure}

\subsection{Training, Validation, and Testing Sets}

One example of these voxel-to-voxel collocations is shown in Figure \ref{fig2}.b. Shown are the selected imager pixels that intersect with profiles from an EarthCARE overpass at 14.875 km and 0.125 km altitude. At the very high viewing zenith angles in Figure \ref{fig2}.b close to the edge of the GOES-19 ABI full disk, there is distinct separation between the two intersections. These two intersections correspond well to A1P1 (a high altitude intersection) and A1P3 (a low altitude intersection) in the illustration in Figure \ref{fig1}.b. Another example in Figure \ref{fig2}.c shows a scene with relatively low viewing angles where these intersections overlap. This is an example similar to P4 in Figure \ref{fig1}.b where the intersections at all altitudes are matched with a single imager pixel and where our voxel-to-voxel approach will yield a similar result to the profile-to-pixel approach.

The overall global distribution of the collocated training, validation, and testing sets are shown in Figure \ref{fig2}.a, d, and e. Several factors contribute to the geographic distribution. The higher temporal and spatial resolution of the ABI and AHI offer more chances for temporally and spatially coincident collocations between the imagers and EarthCARE. Areas with more overlap between multiple geostationary imagers intuitively have an increased number of collocations. We require that all collocations have a time difference of less than 2.5 minutes, accounting for the time it takes the imager scan from the northern to the southern edges of the full disk for ABI/AHI (southern to northern for SEVIRI). Only a single coarsened ACM-CAP estimate is matched to each imager voxel within a maximum horizontal distance of 5 km from the center of the voxel. 

We consider all EarthCARE data between August 1, 2024 to March 10, 2026. GOES-18 ABI, Meteosat-9 and -10 SEVIRI, and Himawari-9 observations span this entire range. GOES-16 was replaced with GOES-19 in the GOES East position on April 7, 2025. We only use GOES-16 data up to March 17, 2025 and use GOES-19 data from April 8th, 2025 onward. The split validation set encompassing early August 2024, and late November 2025 listed in Figure \ref{fig2} ensures that GOES-16 is represented in our validation set. We chose our testing set dates to encompass the operational satellites, and additional training data from December 17th, 2025 onward was added as this work progressed. As a result, GOES-16 is not represented in our EarthCARE testing set, but our ground-based comparison with the MICROBASE products is performed with GOES-16 observations. The validation set was used very minimally for hyperparameter tuning, and the testing set was unused during model development. 

\section{Model Design and Training}
\label{sec:model}

\begin{figure}[t]
    \centering
    \includegraphics[width=16cm]{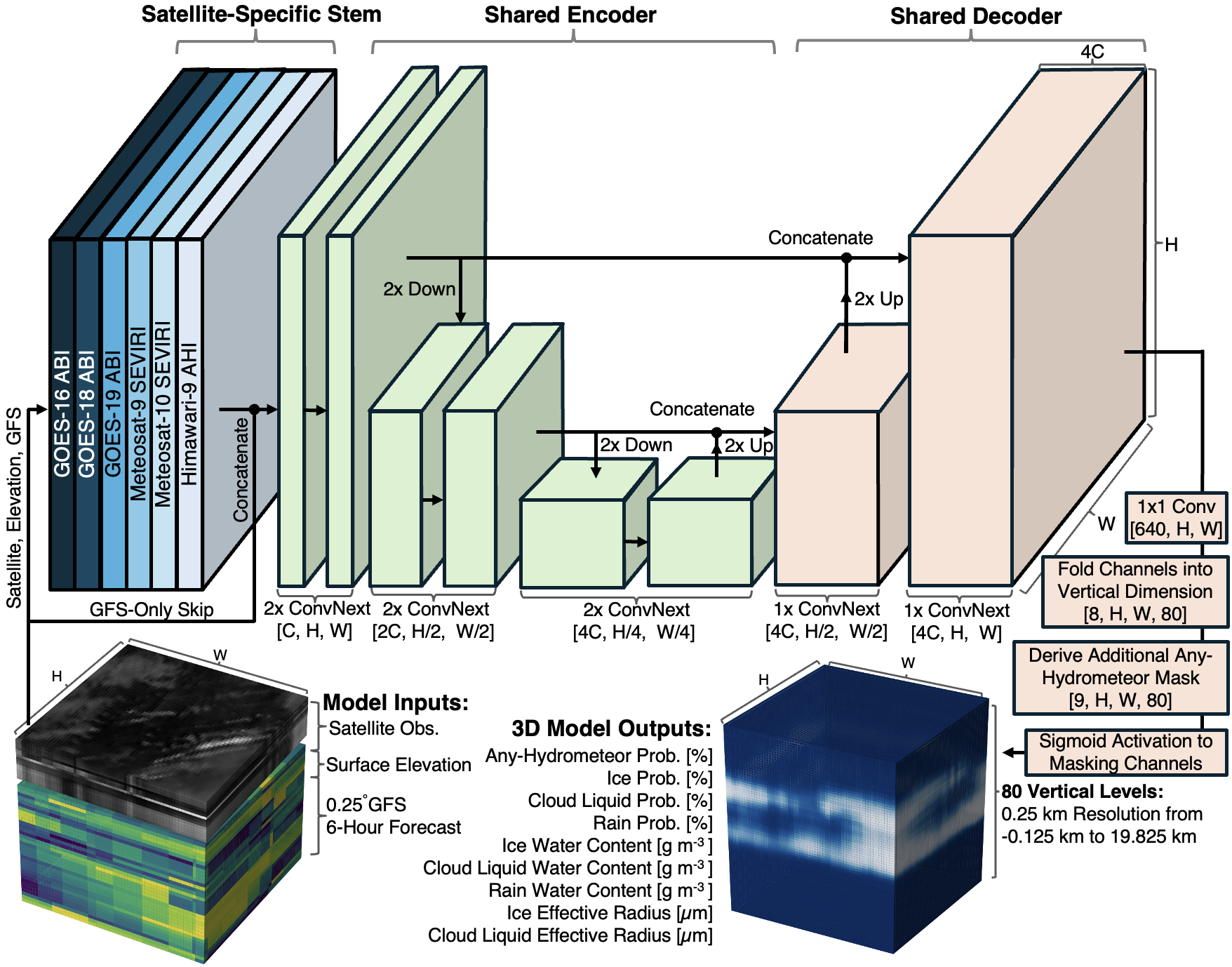}
    \caption{The architecture of our proposed Cloud 3-Dimensional Imager Retrieval (C3DIR) neural network. H, W and C represent the height, width and channel dimension of the output of each block. "2x Down" and "2x Up" represent average pooling and nearest-neighbor interpolated upsampling. The concatenation represents a channel-wise stacking of latent representations. See text for a more detailed discussion of the model.}
    \label{fig3}
\end{figure}

C3DIR can be concisely described as an asymmetrical U-Net composed of 2-D ConvNext \citep{liu2022convnext} blocks with a satellite-specific stem, a shared encoder and decoder, and a depth-to-space operation that yields a 3-D volume with 9 channels (4 masks, 3 water contents, and 2 effective radii). The architecture is shown in Figure \ref{fig3}. The satellite-specific stem is the first component of the model and accepts the inputs consisting of the satellite observations, surface elevation, and a nearest-neighbor resampled GFS forecast. Then, the output activations are pushed through the shared encoder and decoder similar to a typical U-Net. After the final ConvNext block in the shared decoder, the channel dimension is expanded to 640 using a fully-connected layer (or equivalently a 1 x 1 2-D convolution), and the activations are reshaped to a 4-D tensor with 8 channels, 80 vertical levels, and a height and width matching that of the input satellite image. From the first three channels, which represent ice, cloud liquid, and rain logits, we derive an additional any-hydrometer mask by taking the maximum logit of all three classes. The final step is to apply a sigmoid activation to the four masks to yield probabilities. The final output is a 3-D volume with multiple channels. The auxiliary outputs from the NOAA operational products (cloud mask, cloud-top phase, and CTH) are estimated from a small MLP using the activations directly from the last layer of the shared decoder. 

Initially, during training we used the labels from the EarthCARE ACM-CAP product to create a sparsely labeled 3-D volume. We quickly found that this incurred a large amount of unnecessary computational expense due to the large proportion of the volume that was unlabeled. Instead, we treat the output labels as a collection of voxels each with coordinates corresponding to their location in the estimated 3-D volume. Then, those indices are used to select C3DIR voxels to calculate the loss on a much smaller set of input/output pairs. We found this to be a very helpful optimization since typically less than 2.2\% of voxels have labels from EarthCARE.

Below, we discuss the intention behind several of the architectural decisions made when designing C3DIR. 

\subsection{Satellite-Specific Stem}

Each of the three types of passive imaging sensors used in C3DIR have a different selection of VIS, NIR and IR channels. Even the different ABIs aboard GOES-16, -18, and -19 can have differing spectral response functions. Subtle differences in spectral response can have measurable impact on the estimation of cloud properties \citep{smalley2011srf,baum2012srf,meyer2020srf}. Similarly, the geographic areas viewed by each imager are different making it unclear if a model trained on one sensor will generalize well to another of the same or similar design.

One option we tried is to train six separate models for each individual satellite. However, this ignores the possibility that many of the spectral and spatial relationships learned by a model may be similar between sensors. Similarly, it limits our already extremely sparse training data to one-sixth of the total and in our experiments significantly increases the total training time. We also experimented with alternatively using a common subset of roughly similar channels or using feature-wise linear modulation (FiLM; \citealp{perez2017film}) throughout the model to encode satellite-specific information. All of these approaches yielded lower-quality predictions. 

We instead choose to create a module in C3DIR which we call a satellite-specific stem. This module includes a single ConvNext block for each satellite platform we use. In a given sample processed by C3DIR, only one of these blocks is used according to which satellite the sample was obtained from. The input to the selected block within the stem are the observations from that satellite, the GFS fields, and surface elevation. The intention behind this is to allow the model some limited capacity to encode sensor-specific spectral variability or otherwise account for differences between each sensor before the resulting activations are passed to the rest of the model. We expect the greatest benefit to this approach is that is allows a single C3DIR model to train with all collocations from EarthCARE and the six geostationary sensors simultaneously. This allows one sensor's performance to benefit from training data from an entirely different sensor through the use of the shared encoder and decoder described in the next section. This approach shares similarities with EarthNet \citep{vandal2025earthnet}, in which separate vision transformer encoders are used for different satellites are then fed into a shared autoencoder.

\subsection{Asymmetrical Shared Encoder and Decoder}

In contrast to the satellite-specific stem that uses a separate set of parameters for each satellite, the rest of the model which contains the vast majority of learnable parameters, is shared between all satellites. This gives the model the ability to leverage relevant spectral and spatial relationships shared across different satellites and regions during training. 

U-Nets in general, including similar deep learning models used to estimate 3D properties of clouds \citep{wang2023cloud3d,bruning2024cloud3d,jeggle2025cloud3d,amell2024cloud3d,girtsou2025cloud3d}, typically use a symmetrical design with roughly similar dimensions in the encoder and decoder side of the model. In our particular application, the decoder must produce a very high dimensional output amounting to 640 channels. A typical U-Net has a decoder that will gradually contract the channel dimension from the bottom of the U-Net after each up-sampling operation -- usually by one-half each step.

We suggest that channel contraction throughout the decoder can create bottleneck before channel expansion after the decoder in Figure \ref{fig3}. Instead, we keep the channel dimension constant throughout the decoder to lessen the difference in channels between the final ConvNext block and the expansion that immediately follows. Due to the size of the activations produced in the decoder, we only use a single ConvNext block after each up-sampling operation compared to two blocks in the encoder. We found that this asymmetrical design substantially improves the quality and sharpness of the output when compared to symmetrical designs after controlling for similar training throughput.

\subsection{2-D Convolutions with a Depth-to-Space Operation}

In general, tasks involving estimating 3-D volumes often benefit by learning translation-invariant kernels applied across all three dimensions. Consequentially, 3-D convolutions limit the receptive field of a given layer to a fixed height, width, and vertical distance. One of the main challenges in this application is separating out distinct but overlapping cloud layers. These cloud layers can often be several kilometers apart particularly in cases of upper-level thin cirrus overlying low-level cumulus in the tropics. This problem of diagnosing the vertical separation of cloud layers is fundamental to solving this task and we suggest that models benefit from a global receptive field in the vertical dimension throughout all layers. Thus, 3-D convolutions do not seem like an ideal fit for this application. This was our motivation for using 2-D convolutions throughout the model and folding channels dimension into the vertical dimension at the very end. Using 2-D depth-wise separable convolutions followed by linear layers in the ConvNext Blocks, the receptive field in the horizontal dimension is controlled by the size of the kernel, and the channel dimension inherently has a global receptive field. 

When developing this model, we conducted experiments with a combination of 2-D and 3-D ConvNext blocks including using 3-D convolutions throughout the decoder only. The resulting predictions using 3-D convolutions were of substantially lower quality when hyperparameters were tuned to equal training throughput, confirming our final choice of 2-D convolutions.

\subsection{Training Hyperparameters and Loss Function}

C3DIR uses 64 filters in the satellite specific stem and the first block of the encoder. After each down-sampling operation, which is a \(2 \times 2\) average pool, the number of channels is doubled  to a maximum on 256 channels in the bottleneck. The decoder keeps the channel dimension at 256 while up-sampling the activations back to the original spatial resolution. All ConvNext blocks used in the model use a kernel size of \(5 \times 5\), which we found to perform roughly similar to the suggested default of \(7 \times 7\).  We add 12.5\% dropout after the channel expansion in each block, and use a linearly increasing stochastic depth from 0\% at the start of the encoder to 25\% and the end of the decoder. 

Since the satellite-specific stem contains modules that only operate on a subset of samples and our output labels are extremely sparse, we find that batches need to be large and include samples from all satellites. We use a batch size of 1,024 samples randomly selected with replacement and divided among each of the six imaging satellites used. In order to accommodate the large batch size, we use relatively small \(64 \times 64\) image patches during training. Larger image sizes resulted in using less samples per batch and worsened performance of the model significantly. We expect this is due the the increasingly small fraction of labeled voxels with increasing image size and acknowledge that this significantly limits the spatial receptive field of the model. 

C3DIR is trained with a loss function made up of multiple components including labels from both the EarthCARE ACM-CAP product and several imager products outside the EarthCARE ground track which we consider to be an auxiliary loss. Equation \ref{eq:earthcare_voxel_loss} represents the EarthCARE loss (\(\mathcal{L}_{i}^{\mathrm{EC}}\)) calculated for the $i$th voxel where \(\mathcal{K}=\{\mathrm{any\ hydrometer},\mathrm{ice},\mathrm{liquid},\mathrm{rain}\}\), \(\mathcal{S}=\{\mathrm{ice},\mathrm{liquid},\mathrm{rain}\}\), and \(\mathcal{R}=\{\mathrm{ice},\mathrm{liquid}\}\) denote the mask, water-content, and effective-radius output sets, respectively. Here, $\hat{p}_{i,k}$ and $p_{i,k}$ denote the predicted and labeled probabilities for species $k$ at voxel $i$, respectively. Similarly, $\hat{q}_{i,k}$ and $q_{i,k}$ denote the predicted and labeled water content, and $\hat{r}_{i,k}$ and $r_{i,k}$ denote the predicted and labeled effective radius.

\begin{equation}
\mathcal{L}_{i}^{\mathrm{EC}}
=
\underbrace{
\gamma_{\mathrm{mask}}
\sum_{k \in \mathcal{K}}
\omega_{i,k}^{\mathrm{mask}}
\,
\mathrm{BCE}\!\left(\hat{p}_{i,k}, p_{i,k}\right)
}_{\text{EarthCARE Masks}}
+
\underbrace{
\sum_{k \in \mathcal{S}}
\left(
\hat{q}_{i,k} - q_{i,k}
\right)^2
}_{\text{EarthCARE Water Contents}}
+
\underbrace{
\sum_{k \in \mathcal{R}}
\left(
\hat{r}_{i,k} - r_{i,k}
\right)^2
}_{\text{EarthCARE Effective Radii}} .
\label{eq:earthcare_voxel_loss}
\end{equation}

\begin{equation}
\omega_{i,k}^{\mathrm{mask}}
=
\begin{cases}
1, & p_{i,k}=1, \\
(1-\lambda)
+
\lambda \,
\max\left(0.1, F_{z_i,k}\right), & p_{i,k}=0 .
\end{cases}
\label{eq:mask_weights}
\end{equation}

For the masking variables we use binary cross-entropy (BCE) weighted by $\omega_{i,k}^{\mathrm{mask}}$ (Eq. \ref{eq:mask_weights}), which is based on the fractional occurrence of hydrometeors at a given altitude and is calculated separately for each mask. This fractional occurrence is represented by $F_{z_i,k}$ where $z_{i}$ is the altitude of voxel $i$ belonging to species $k \in \mathcal{K}$. For each of the four masking variables, we assign a weight of 1 indicating the presence of that species of water, but we decrease the weight of clear-sky voxels down to a minimum value of 0.1. This weighting is applied to penalize C3DIR more heavily for voxels with hydrometeors in them, and reduce the influence of the very large number of hydrometeor-free voxels. However, we found that this weighting over-corrected and too strongly incentivized the model to predict the positive class. We added an additional parameter, $\lambda$, which we set to 0.25 and moderates the strength of this weighting. Lastly, we reduce the overall magnitude of the masking term by multiplying it by $\gamma_{\mathrm{mask}}$ which we also set to 0.25. No altitude or species dependent weighting is applied to the water contents or effective radii since clear-sky voxels have no impact on those components.

The auxiliary component of the loss function is represented by $\mathcal{L}_{j}^{aux}$ for the $j$th labeled pixel in Eq. \ref{eq:aux_pixel_loss}. This term combines the BCE loss for the auxiliary cloud mask and cloud-top phase labels with mean squared error for CTH. Here, $\hat{p}_{j,\mathrm{cm}}$ and $p_{j,\mathrm{cm}}$ are the predicted and labeled cloud-mask for pixel $j$, while $\hat{p}_{j,\mathrm{phase}}$ and $p_{j,\mathrm{phase}}$ are the predicted and labeled cloud-top phase. $\hat{h}_{j}$ and $h_{j}$ denote the predicted and labeled CTH for the same pixel. As mentioned previously, we found that adding this auxiliary term improved model performance on the validation set. We expect there may be further benefit in weighting this term more or less, but we did not experiment with this during development of C3DIR.

The total training loss used to train C3DIR is given in Equation \ref{eq:total_loss}. The first term is the average over all voxels and the second term is the average over all pixels with auxiliary labels.

\begin{equation}
\mathcal{L}_{j}^{aux}
=
\underbrace{
\mathrm{BCE}\!\left(\hat{p}_{j,\mathrm{cm}}, p_{j,\mathrm{cm}}\right)
}_{\text{Aux. Cloud Mask}}
+
\underbrace{
\mathrm{BCE}\!\left(\hat{p}_{j,\mathrm{phase}}, p_{j,\mathrm{phase}}\right)
}_{\text{Aux. Cloud-top Phase}}
+
\underbrace{
\left(
\hat{h}_{j} - h_{j}
\right)^2
}_{\text{Aux. Cloud-top Height}} .
\label{eq:aux_pixel_loss}
\end{equation}

\begin{equation}
\mathcal{L}
=
\frac{1}{n_{\mathrm{v}}}
\sum_{i=1}^{n_{\mathrm{v}}}
\mathcal{L}_{i}^{\mathrm{EC}}
+
\frac{1}{n_{\mathrm{p}}}
\sum_{j=1}^{n_{\mathrm{p}}}
\mathcal{L}_{j}^{\mathrm{aux}} .
\label{eq:total_loss}
\end{equation}

C3DIR is trained for 200,000 forward and backward passes using a cosine annealing learning rate schedule starting at a learning rate of $1 \times 10^{-3}$ and ending with $1 \times 10^{-6}$. Batches are randomly augmented with horizontal and vertical flips with 50\% probability.  The AdamW \citep{kingma2017adam,loschilov2019adamw} optimizer is used with weight decay set to 0.1. It takes roughly 55 hours to train C3DIR using four NVIDIA RTX A6000 GPUs using data parallelism.

\subsection{Input and Output Standardization}

All of the VIS, NIR, and IR channels as well as the surface elevation are standardized using the mean and standard deviation of each channel. Separate sets of means and standard deviations are used for each satellite to account for different spectral response or geographic area viewed. The GFS information is standardized between 0 and 1 using the minimum and maximum values for each channel. The same set of GFS minimum and maximum values are used across all satellites. All input standardization statistics are determined from the training dataset.

No standardization is applied to the masking output fields derived from the EarthCARE ACM-CAP product given that they are already binary. Water content spans several orders of magnitude in the ACM-CAP product which poses some difficulty in terms of scaling. We choose to use a logarithmic scaling using Eq. \ref{eq:logscale} with a scaling constant, $s = 5 \times 10^{-5}$, unscaled water content $w$, and a scaled water content $\tilde{w}$.

\begin{equation}
\tilde{w} = \log_{10}\left(1 + \frac{w}{s}\right)
\label{eq:logscale}
\end{equation}

The auxiliary NOAA operational cloud mask and cloud phase already exist between 0 and 1 and require no scaling. The remaining auxiliary outputs are scaled linearly and intended to keep the most values between 0 and 1. CTH, expressed in meters, is divided by 12,000 m. The effective radii are both expressed in µm. Ice effective radii larger than 140 µm are truncated and all values are divided by 20 µm. Liquid effective radii larger than 35 µm are truncated and all values are divided by 5 µm.

\section{Results and Model Performance Analysis}
\label{sec:results}

All analyses performed here use the held-out test set depicted in Figure \ref{fig2} and six months of the MICROBASE retrieval using the ground-based instrumentation at the SGP-ARM site (referred to hereafter as the SGP-ARM products). The primary testing set used is the coarsened EarthCARE ACM-CAP product similarly used to train C3DIR, and the comparisons between the SGP-ARM and C3DIR are shown later in Section \ref{sec:ground_based}. First, we show two example case studies of C3DIR predictions with collocated EarthCARE ground tracks.


\subsection{Qualitative case studies}

\subsubsection{GOES-19 example}

\begin{figure}[t]
    \centering
    \includegraphics[width=15.5cm]{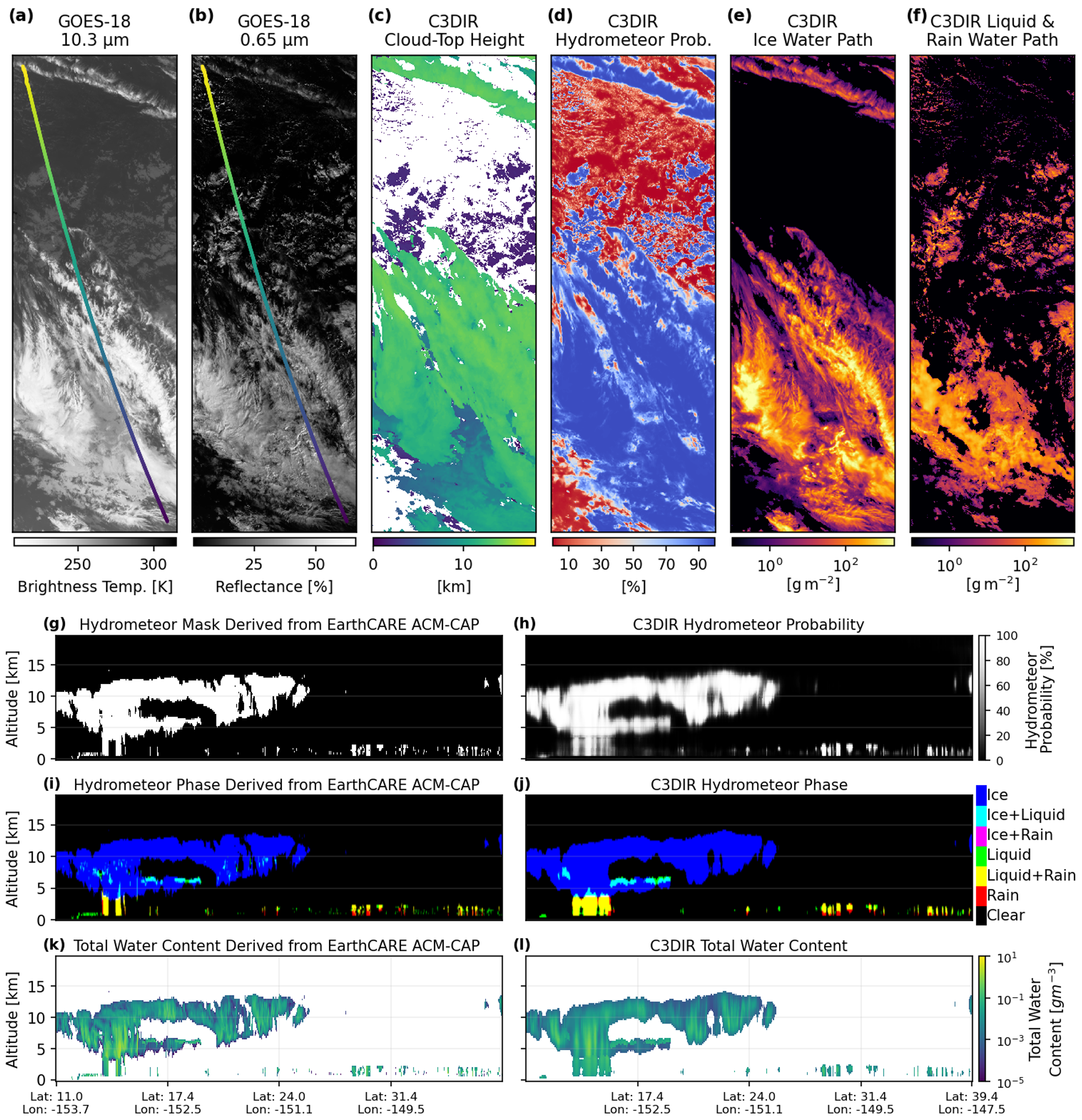}
    \caption{A sample GOES-18 ABI scene with a collocated EarthCARE overpass on December 1 00:10 UTC centered on \(21^\circ\)N latitude and \(151^\circ\)W longitude. (a) and (b) are the observed \(10.3~\mu\mathrm{m}\) and \(0.65~\mu\mathrm{m}\) channels from ABI where the line crossing the image represents the start (purple) and the end (yellow) of the collocated EarthCARE ground track. (c)-(f) represent products derived from the estimated 3-D volume from C3DIR (see text for details). (g), (i), and (k) represent collocated cross sections derived from the EarthCARE ACM-CAP product. (h), (j) and (l) are the C3DIR estimates corresponding to those cross sections. Note the the beginning (left-hand side) of the vertical cross sections correspond to the purple ground track in (a) and (b) and the end (right-hand side) corresponds to the yellow.}
    \label{fig4}
\end{figure}

Figure \ref{fig4} illustrates a variety of products derived from the 3-D volumes estimated from C3DIR and collocated estimates from ACM-CAP. Figure \ref{fig4}.a and \ref{fig4}.b show the imager perspective of these clouds with the EarthCARE ground track overlaid on top. The EarthCARE ACM-CAP product determines this to be a multi-layer scene with upper-level ice clouds overlaying a mid-level liquid-topped mixed-phased cloud layer. The ACM-CAP product also detects some amount of precipitation extending down to the surface in the southern portion of the image. A large amount of the water content in the ACM-CAP profiles appears to be associated with these precipitating clouds, but with some finer scale variations in ice water content among the upper-level ice clouds. There is also a thin liquid layer layer with roughly $10^{-2}$ to $10^{-1}$ \(\mathrm{g m^{-3}}\) water content. There are also finer-scale low-level clouds placed sporadically across the northern portion of the image and the vertical cross section.

Overall, the vertical cross section from C3DIR (Figure \ref{fig4}.h,j,l) matches this characterization well with a few caveats. The vertical extent of the upper-level ice cloud layer matches very well with ACM-CAP, but with some smoothing in upper-level ice cloud boundaries. The presence of a second-layer between 5 to 6 km is identified in C3DIR predictions including some amount of cloud liquid embedded in the precipitating and liquid-topped clouds seen in ACM-CAP. The C3DIR liquid layer appears to be more geometrically thick. C3DIR correctly identifies the presence of precipitation, but overestimates the horizontal extent of the precipitating clouds, appearing to double the precipitating area. Smaller low-level clouds in the north of the image are mostly identified, but some appear to be smoothed over. In non-precipitating profiles, the total water content (TWC) aligns closely with the ACM-CAP product, but with a significantly smoother appearance lacking much of the finer detail typically seen from active profiling instruments.

Despite some of the above mentioned issues, the overall scene is well-represented by C3DIR. The CTH estimates derived from C3DIR (Figure \ref{fig4}.c) illustrate cold cirrus that extend over the central portion of this scene with a few gaps allowing views of mid-level cloud-tops. The C3DIR 2-D hydrometeor probabilities in Fig \ref{fig4}.d are derived by finding the highest predicted voxel probability in a given pixel. This mask appears to agree well with what is seen in the \(10.3~\mu\mathrm{m}\) and \(0.65~\mu\mathrm{m}\) imagery, albeit with some smoothing. Probabilities close to 50\% typically occur among optically thin clouds and cloud edges where C3DIR is more uncertain. Cloud edges can often be more optically thin due to entrainment processes which reduce the amount of condensate and may contribute higher uncertainty in the presence of hydrometeors along with sub-pixel heterogeneity. We also expect that the smoothness in the cloud probabilities, to some degree, illustrates issues associated with larger temporal separations between the active and passive instruments. Large time differences allow for clouds to advect out of view or into view in a given area meaning labels are typically noisier around cloud edges and clouds with small horizontal scale. 

Ice water path (IWP) in Figure \ref{fig4}.e and the sum total of liquid water path (LWP) rain water path (RWP) in Figure \ref{fig4}.f are difficult to assess from qualitative means alone. Intuitively, we observe higher IWP and liquid and rain water path corresponding to colder and brighter clouds in the IR and VIS imagery, respectively. We also observe elevated LWP and RWP associated with the precipitating cloud profiles and the liquid-topped mixed-phase clouds. LWP and RWP are smoother in their appearance likely due to the lack of direct spectral information from GOES-18 ABI beneath optically thick cloud layers. This in contrast to the upper-level ice clouds which are closer to where the IR and VIS weighting functions are more likely to peak. Although C3DIR produces RWP estimates trained to match the ACM-CAP product, quantitatively accurate RWP retrieval from IR and VIS imager measurements alone is not expected to be as well constrained and we consider it to beyond the scope of C3DIR.

\subsubsection{Himawari-9 example}

\begin{figure}[t]
    \centering
    \includegraphics[width=15.5cm]{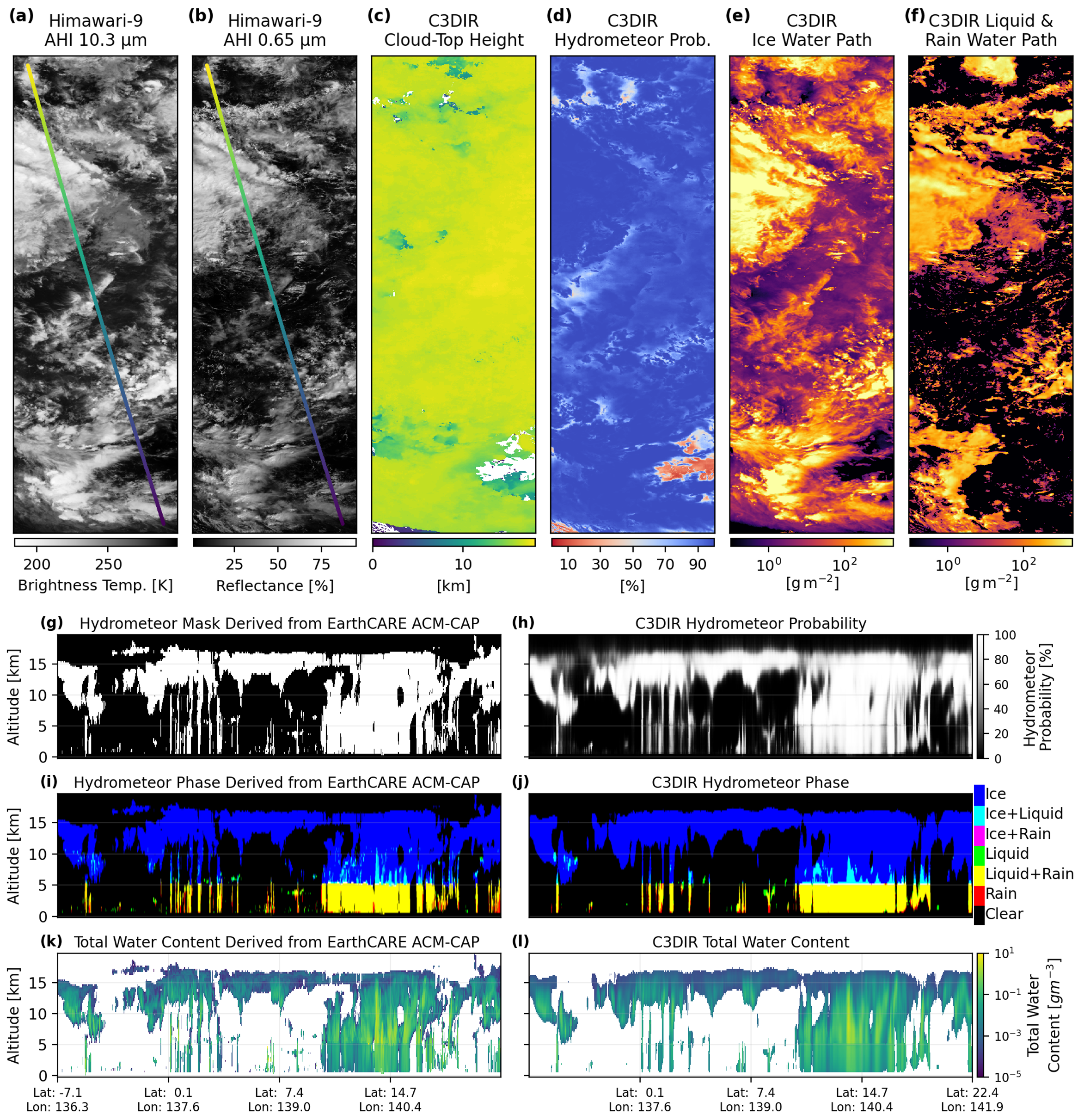}
    \caption{A sample Himawari-9 AHI scene with a collocated EarthCARE overpass on December 3 04:40 UTC centered on 7 N latitude and 138 E longitude. (a) and (b) are the observed \(10.3~\mu\mathrm{m}\) and \(0.65~\mu\mathrm{m}\) channels from ABI where the line crossing the image represents the start (purple) and the end (yellow) of the collocated EarthCARE ground track. (c)-(f) represent products derived from the estimated 3-D volume from C3DIR (see text for details). (g), (i), and (k) represent collocated cross sections derived from the EarthCARE ACM-CAP product. (h), (j) and (l) are the C3DIR estimates corresponding to those cross sections. Note the the beginning (left-hand side) of the vertical cross sections correspond to the purple ground track in (a) and (b) and the end (right-hand side) corresponds to the yellow.}
    \label{fig5}
\end{figure}

Figure \ref{fig5} illustrates a scene from Himawari-9 AHI. This is another multilayer scene, but with several convective clouds and a much broader area of precipitation in the the northern half of the image. The ACM-CAP product detects a broad expanse of thin cirrus throughout this region. In the vertical cross section, there is a strong co-occurrence of both cloud liquid and rain in the ACM-CAP product. Most of the TWC is contained in the broader precipitating region and only very small amounts are contained in upper level ice clouds approaching the $10^{-5}$ \(\mathrm{g m^{-3}}\) threshold we use to discriminated cloudy from clear voxels for the purposes of C3DIR.

Similar to the previous example, C3DIR appears to reliably characterize the vertical cross section with a few caveats. C3DIR correctly identifies the upper-level ice clouds, but misses a few finer-scale cloud-free gaps and overestimates the geometric thickness of the layer in a few locations. C3DIR correctly identifies the presence of liquid embedded within the ice cloud, but occasionally overestimates the thickness of those liquid layers. The model also correctly identifies the presence of rain depicted in the ACM-CAP product and the transition from cloud ice and liquid to liquid and rain. Some of the convective towers identified in the ACM-CAP product are missing from C3DIR underneath the more optically thick portions of the upper-level ice clouds. The TWC in the in the vertical cross sections match surprisingly well between C3DIR and EarthCARE given the sometimes limited information present in passive observations. The vertical distribution and amounts in the upper-level ice clouds and even in the deeper precipitating region show relatively good agreement.

C3DIR CTH depicts the upper-level cirrus as covering the almost the entire region in this example with somewhat high confidence in the predicted hydrometeor probabilities. This is certainly plausible given what is observed in the ACM-CAP cross-section. C3DIR attributes very small IWP (roughly 1 to 3 \(\mathrm{g m^{-2}})\) to these profiles that otherwise do not contain any lower-level or convective clouds. This serves as a useful example of the benefit of characterizing multiple cloud layers since the relatively thin cirrus may or may not have impacts on downstream applications. By explicitly characterizing multiple layers and their properties throughout the vertical profile, we expect that C3DIR's outputs can be more easily adapted to downstream products since they are less constrained by development-time decisions regarding which cloud layers should be represented.

\subsection{Voxel-level evaluation}

Next we perform a more quantitative evaluation of the quality of C3DIR's output continuing to use the EarthCARE ACM-CAP product as the baseline. Here, we focus primarily on the voxel-level performance. We again note that imager observations are the combined result of radiative contributions of varying levels of the atmosphere, cloud layers and surface. As a result, the performance of detecting a cloud or estimating water content at a specific altitude cannot easily be separated from the task of placing the cloud layer in the vertical dimension. For example, the hydrometeor detection skill in a given voxel cannot be entirely separated from the skill C3DIR has in height assignment of a given cloud layer. This is most problematic for geometrically thin clouds, where placing a 500 m thick cloud layer 500 m higher in the vertical yields both voxel detection and water content errors.
\begin{figure}[t]
    \centering
    \includegraphics[width=16cm]{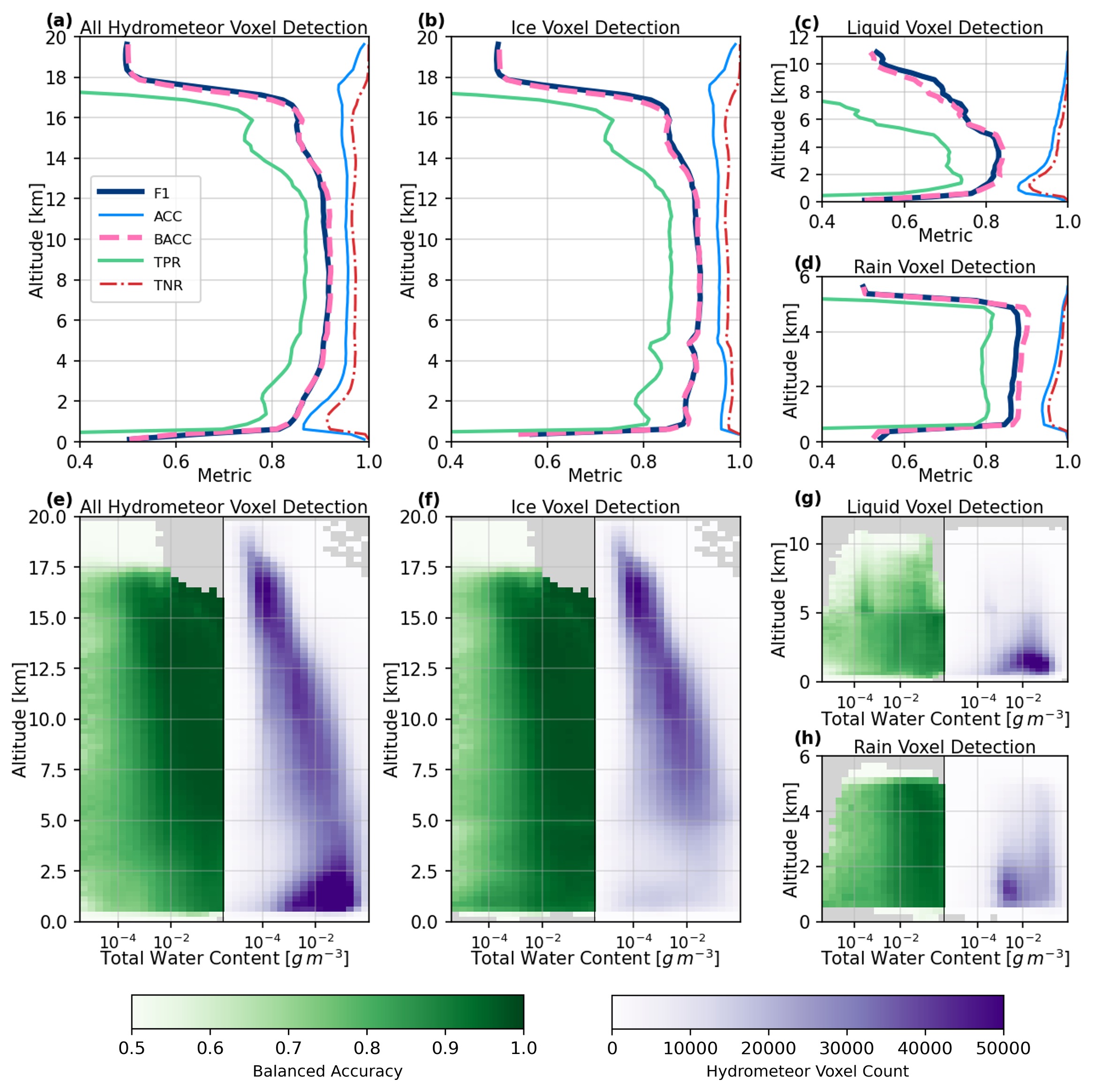}
    \caption{Evaluation of voxel hydrometeor detection. (a)-(d) show the macro F$_{1}$ score, ACC, BACC, TPR, and TNR for detecting the presence of each class as a function of altitude. (e)-(h) show the BACC as a function of altitude and the TWC. Note that the green colorbar corresponds to the BACC in (e)-(h) and the purple colorbar corresponds to the number of voxels belonging to the positive class for each combination of altitude and TWC.}
    \label{fig6}
\end{figure}

\subsubsection{Hydrometeor Detection Performance}

We start with an evaluation of how well C3DIR detects any hydrometeor, ice, liquid, and rain voxels. For the purposes of this evaluation, we focus on the ability of C3DIR to discriminate the presence from absence of these categories (e.g., ice or ice-free voxels). This is in contrast to an evaluation of identifying, for example, ice-only from mixed-phase voxels which we leave to our future work. Figure \ref{fig6} shows several metrics for detecting these categories as a function of altitude and TWC. We examine the F$_{1}$-score (Eq. \ref{eq:f1}), accuracy (ACC; Eq. \ref{eq:acc}),  True Positive Rate (TPR; Eq. \ref{eq:tpr}), True Negative Rate (TNR; Eq. \ref{eq:tnr}), and balanced accuracy (BACC; Eq. \ref{eq:bacc}). In these equations, TP represents the number of true positives, TN are true negatives, FP are false positives and FN are false negatives. Cloud or hydrometeor detection can be a heavily imbalanced classification problem when looking at specific altitude levels, so we focus this analysis primarily on BACC which accounts for this. Note that we use the macro-F$_{1}$ score in Figure \ref{fig6} which is the average of Eq. \ref{eq:f1} when both "cloud" and "cloud-free" are used as the positive class.

\begin{equation}
\text{F}_1 = \frac{2\mathrm{TP}}{2\mathrm{TP} + \mathrm{FP} + \mathrm{FN}}
\label{eq:f1}
\end{equation}

\begin{equation}
\mathrm{ACC} = \frac{\mathrm{TP} + \mathrm{TN}}{\mathrm{TP} + \mathrm{TN} + \mathrm{FP} + \mathrm{FN}}
\label{eq:acc}
\end{equation}

\begin{equation}
\mathrm{TPR} = \frac{\mathrm{TP}}{\mathrm{TP} + \mathrm{FN}}
\label{eq:tpr}
\end{equation}

\begin{equation}
\mathrm{TNR} = \frac{\mathrm{TN}}{\mathrm{TN} + \mathrm{FP}}
\label{eq:tnr}
\end{equation}

\begin{equation}
\mathrm{BACC} = \frac{1}{2} \left( \mathrm{TPR} + \mathrm{TNR} \right)
\label{eq:bacc}
\end{equation}

Across all four categories in in Figure \ref{fig6}.a-d, we note that TPR is much lower than TNR. This is partially due to the way we have constructed the positive class in the labeled dataset which consists of any voxel with a water content exceeding $1 \times 10^{-5}$ \(\mathrm{g m^{-3}}\). This is an extremely small value that the ACM-CAP product struggles to characterize well \citep{mason2023acmcap}. Thus, we are training and evaluating C3DIR to identify exceptionally tenuous clouds that are not easily identified particularly from passive remote sensing instruments. The low TPR values imply that there is a substantial fraction of clouds that are not identified or misplaced in the vertical. TNR, on the other hand, is typically greater than 95\% meaning that hydrometeor-free voxels, which constitute the majority class, are often predicted correctly. This difference in TNR and TPR results in C3DIR predicting fewer cloudy voxels compared to EarthCARE ACM-CAP. Since hydrometeor-free voxels represent the majority class, ACC tracks closely with TNR. BACC and the macro F$_{1}$-score both account for class imbalance and track very closely together across all altitudes and water species.

Overall performance for the ice and all-hydrometeor masks is best between 17 km and 1 km. Above 17-km the positive class becomes very sparse and C3DIR tends to predict it infrequently which is evidenced by the extremely low TPR at high altitudes. The clouds that occur above 17 km typically have a TWC less than $1\times10^{-4}$ \(\mathrm{g m^{-3}}\) and are likely optically thin. Below 750 m, the ACM-CAP product rarely indicates the presence of cloud. C3DIR appears to inherit this behavior and rarely extends any of the water species down to the surface. We expect that this is a strong limitation of using space-borne active sensors as the labeled dataset where radar returns very close to the surface are screened out due to possible contamination from surface clutter \citep{schulte2023cloudsat}. Aside from these two issues, performance of the all-hydrometeor mask and the ice mask are relatively consistent at all altitudes with a slight drop in performance above 12-km.

When BACC is examined as a function of TWC (Figure \ref{fig6}.e), we note that BACC is not strongly related to altitude, but is more so a function of how much water is in the voxel. Liquid voxel detection is best between 2-6 km, but BACC, F$_{1}$-score, and TPR decrease steadily above 6 km. We expect this is due C3DIR misclassifying liquid voxels embedded within large ice-clouds as ice-only. There is a steep drop in performance above 6 km where it is possible liquid clouds are being misclassified as ice-only where ACM-CAP has identified mixed-phase clouds. Indeed, in the cases shown in Figures \ref{fig4} and \ref{fig5} we note many finer-scale mixed-phase clouds in the EarthCARE ACM-CAP cross section that are not identified in the C3DIR output. Liquid clouds also, by nature of being lower in the atmosphere, may be obscured by upper-level ice clouds which complicates their detection.

Raining voxel detection performs slightly better than liquid voxel detection between 1-5 km, which may seem to be an unintuitive result from a passive imaging instrument. We believe this difference is again due to the problem of height assignment. Liquid cloud layers are geometrically thin meaning small displacements in the vertical hurt performance. In vertical profiles with precipitation where rain is detected, it is likely to extend from within the cloud down to the surface. Therefore, the primary challenge that C3DIR needs to solve is identifying the presence of rain in the profile, and we expect that the vertical placement of those raining voxels is a secondary and simpler task once such profiles are identified. 

\subsubsection{Water Content Quantification}

\begin{figure}[t]
    \centering
    \includegraphics[width=16cm]{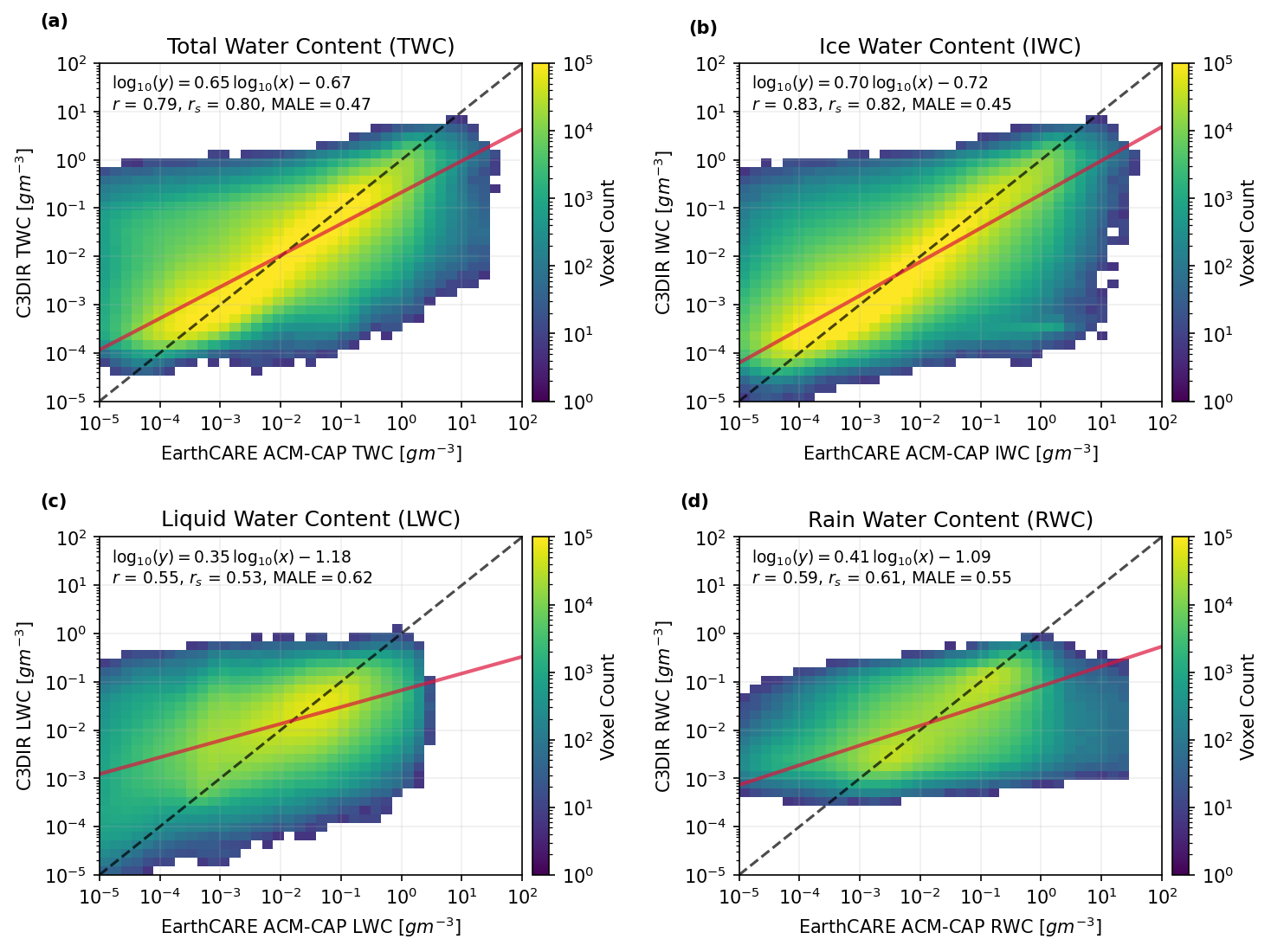}
    \caption{Evaluation of C3DIR estimates of (a) TWC, (b) IWC, (c) LWC, and (d) RWC. Noted inside each plot is the equation of a linear fit between the ACM-CAP water contents and the predicted values. Also shown are the Pearson correlation coefficient ($r$), Spearman correlation coefficient ($r_{s}$), and the MALE.}
    \label{fig7}
\end{figure}

Next we move to an evaluation of how well C3DIR estimates the total, ice, cloud liquid, and rain water contents in the ACM-CAP product. This is a particularly challenging task as passive imaging instruments often lack sensitivity to within-cloud variation of water content. Since differences in water content often span orders of magnitude, we use the mean absolute log error (MALE; Eq. \ref{eq:male}) as one of our primary metrics where $\hat{y}_{i}$ and $y_{i}$ are the predicted and labeled quantities for the $i$th voxel. 

\begin{equation}
\mathrm{MALE}
=
\frac{1}{N}
\sum_{i=1}^{N}
\left|
\log_{10}(\hat{y}_i)
-
\log_{10}(y_i)
\right|
\label{eq:male}
\end{equation}

Figure \ref{fig7} shows the relationship between the C3DIR-predicted water contents and those provided by ACM-CAP. C3DIR appears to have a moderate positive correspondence in TWC (Figure \ref{fig7}.a) with a majority of estimates centered on the 1:1 line. However, we note that in many cases C3DIR overestimates low TWC values and underestimates high TWC. This is a very common characteristic of AI/ML models and we expect this is related to the spatial smoothing seen in Figures \ref{fig4}.l and \ref{fig5}.l. The IWC scatter plots in Figure \ref{fig7}.b appear very similar to the the TWC comparison which is unsurprising given that ice voxels are much more common than liquid and rain voxels. 

Liquid and rain water content estimates from C3DIR do not match the ACM-CAP product as well as ice. LWC estimates (Figure \ref{fig7}.c) have lower correlation coefficients and higher MALE compared to IWC. The overall pattern looks similar, but with significantly higher spread about the 1:1 line, and a stronger C3DIR high bias for low ACM-CAP LWC. RWC estimates are similarly poor. While there is still a moderate positive relationship between C3DIR RWC and ACM-CAP RWC indicated by a Pearson correlation coefficient of 0.59, there is a similarly large degree of spread about the 1:1 line and bias present in the C3DIR estimates at extreme high and low ACM-CAP values. For rain specifically, we note that the ACM-CAP product has subtle, but distinct peaks in the RWC distribution at roughly $5\times 10^{-3}$ and $1\times10^{-1}$  \(\mathrm{g m^{-3}}\). These peaks correspond to RWC resulting from warm-rain and cold-rain processes. We expect that the limited skill that C3DIR has in estimating the ACM-CAP RWC likely comes from identifying the whether the rain present in the profile is resulting from how warm-rain or cold-rain processes are depicted in the ACM-CAP products and not from direct physical sensitivity to RWC in the passive observations. We expect that further evaluation work is needed to determine what information C3DIR uses to quantify RWC, and further validation studies are needed to assess the performance of the ACM-CAP product in characterizing precipitating liquid clouds. 

 \subsection{Profile-level Evaluation}

Next, we focus our analysis on how well C3DIR characterizes properties of the vertical profile. Similar to the cross-sections shown in Figures \ref{fig4} and \ref{fig5}, we construct vertical profiles from C3DIR that are normal to the surface by sampling voxels from multiple imager pixels at the corresponding altitudes. The result is a profile of the atmosphere that matches the view seen from EarthCARE. Thus, the problem of misaligned views between the instruments is accounted for in these comparisons between C3DIR and EarthCARE. We perform comparisons with the NOAA operational products where appropriate. Since the operational products do not produce vertical profiles of the atmosphere, a similar resampling cannot be performed. Instead, we use the CTH parallax correction approach to find matching estimates from the operational products. Thus, the differences in performance seen here include differences in the estimates themselves and the ability of C3DIR to construct truly vertical profiles from its 3-D output that more closely align with nadir-pointing active sensors.

\subsubsection{Cloud-Top and Cloud-Base Height}

Estimating cloud boundaries such as CTH and cloud-base height (CBH) is a common application of passive imaging instruments. Using the 3-D output from C3DIR, we can derive more traditional cloud products CTH (as seen in Figures \ref{fig4}.c and \ref{fig5}.c) and CBH by finding the altitude of the lowest and highest point in the profile that contains a hydrometeor. Note that CBH can be defined in different ways for multi-layered scenes and may or may not include precipitation. In the analysis shown here, we choose CBH to be the lowest altitude that any hydrometeor is detected. For the purposes of these evaluations, we calculate geometric thickness based on the total vertical distance occupied by hydrometeors (not including clear-sky gaps) regardless of the number of distinct layers in the profiles. 

\begin{figure}[t]
    \centering
    \includegraphics[width=16cm]{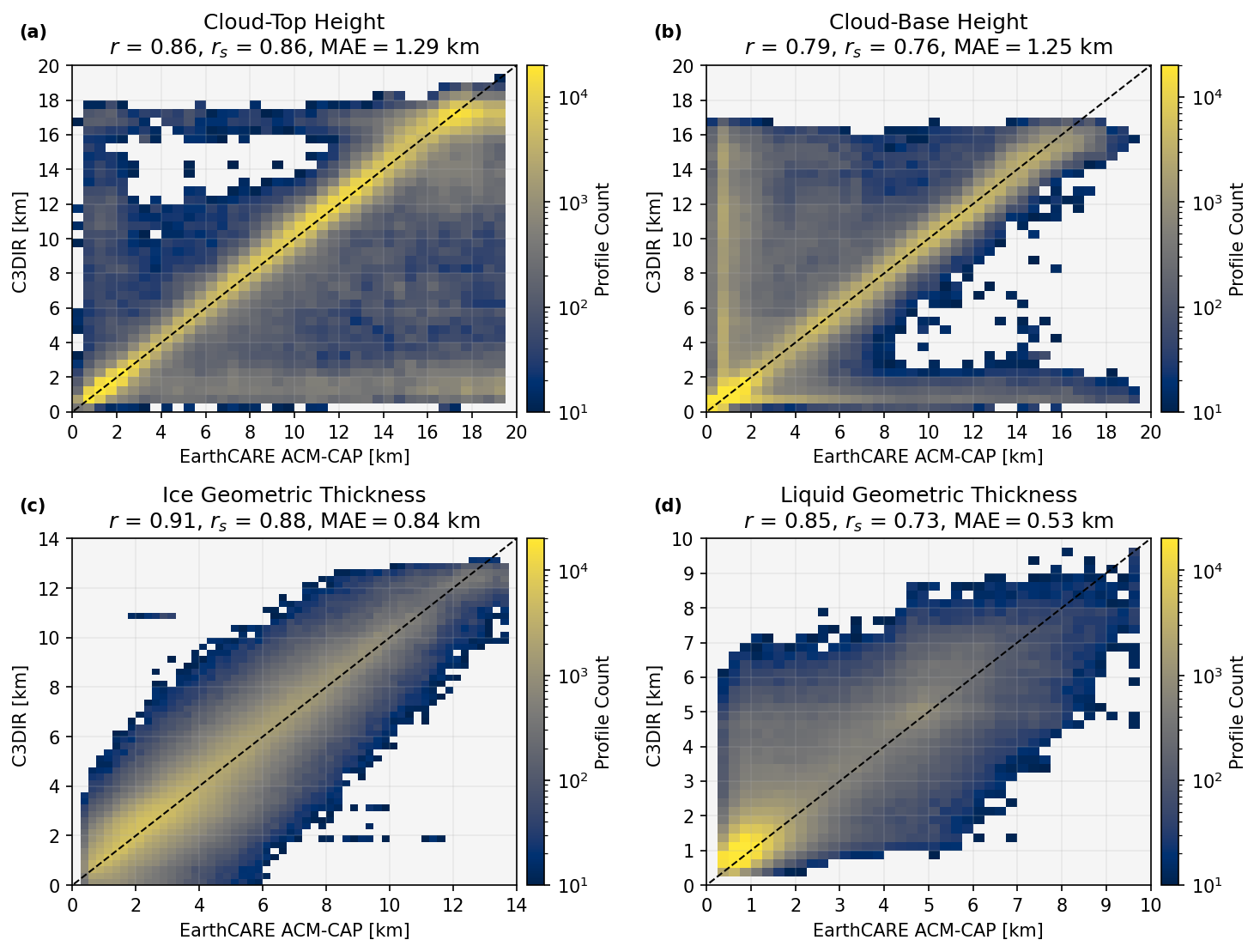}
    \caption{Comparison of C3DIR (a) CTH, (b) CBH, (c) ice geometric thickness, and (d) liquid geometric thickness. Note that the geometric thicknesses shown here include the sum total of vertical distance where each species are present regardless of the number of distinct cloud layers. Also shown are the Pearson correlation coefficient ($r$), Spearman correlation coefficient ($r_{s}$), and the mean absolute error (MAE).}
    \label{fig8}
\end{figure}

Figure \ref{fig8} shows the comparison of C3DIR CTH, CBH, as well as ice and cloud liquid geometric thickness. Overall, there is very good agreement in CTH between C3DIR and the ACM-CAP product. The largest errors from C3DIR typically occur where C3DIR estimates a CTH lower than 2 km. Based on the results seen in the voxel hydrometeor detection analysis, we expect these to be multi-layer scenes where EarthCARE sees a high-altitude optically thin cloud or cloud edge above a lower level opaque cloud. In such cases C3DIR might miss the high-altitude cloud, but reliably detect a cloud present a the lower levels. C3DIR had relatively few cases where it overestimates CTH relative to ACM-CAP.

CBH takes on a somewhat smaller range of values because it is bounded between CTH and the surface elevation. Similar to CTH, most samples are along the 1:1 line, but there are some exceptionally large errors. We expect that the high C3DIR overestimates are again multi-layer cases where a low-level cloud is underneath an upper-level opaque cloud missed by C3DIR. There is also a large subset of cases where C3DIR estimates a CBH lower than 2 km, but ACM-CAP places the CBH anywhere from 0 km to 18 km. We expect these cases represent false low-cloud detection from C3DIR, or false identification of precipitation reaching the surface. 

Total ice cloud geometric thickness appears to be well represented by C3DIR with a Pearson correlation coefficient of 0.91 and MAE of 0.84 km. Cloud liquid geometric thickness has a somewhat lower correlation coefficient of 0.85 and an MAE of 0.52. A large proportion of liquid clouds are geometrically thin with thicknesses frequently between 0.5 km and 2 km. We note that liquid cloud geometric thickness tends to be slightly overestimated. This is apparent in the case study shown in Figure \ref{fig4}.j where the liquid-topped mixed phase cloud spans a greater vertical distance in C3DIR compared to EarthCARE.

\begin{figure}[t]
    \centering
    \includegraphics[width=16cm]{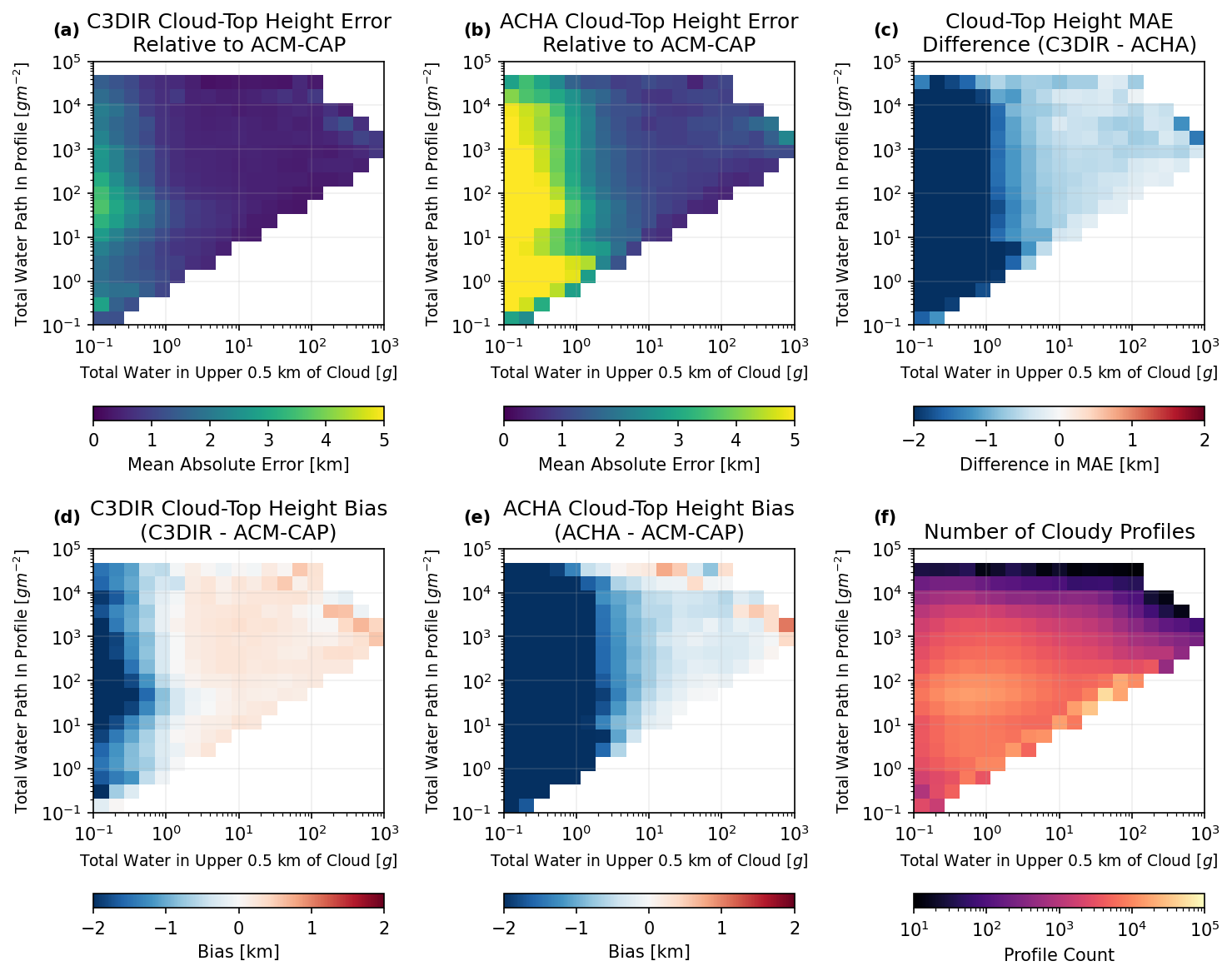}
    \caption{Evaluation of C3DIR and and ACHA CTH performance relative to ACM-CAP. (a) and (b) shown the MAE of C3DIR and ACHA as a function of the total amount of water in the profile and the total amount of water in the uppermost 0.5 km of the cloud. (c) shows the difference in MAE between C3DIR and ACHA. (d) and (e) show the mean bias relative to ACM-CAP. (f) shows the number of cloudy profiles. }
    \label{fig9}
\end{figure}

Given that C3DIR uses a 3-D retrieval framework, it is worth comparing with more traditional approaches to examine potential differences in resulting cloud products. However, one of the primary challenges when comparing cloud property algorithms centers around a given approach's effective definition of what constitutes a cloud, the typical downstream applications an algorithm is tuned for, and sometimes subtle differences in definition of a particular cloud property. Here, we compare the CTH derived from C3DIR to CTH estimated from ACHA processed within the CLAVR-x retrieval system. C3DIR is trained to identify cloudy voxels with extremely low water content by passive measurement standards. We essentially train C3DIR to identify a CTH closer to what would be identified by a lidar. ACHA must support a variety of downstream applications \citep{heidinger2020achaatbd} including imager-derived winds which benefit from cloud-top heights deeper within optically thin cloud layers compared to what would be typically given by a lidar \citep{dimichele2013cthlidar}.

In order to address these potential differences between these two algorithms, we perform an analysis that quantifies differences in error with respect to ACM-CAP based on how tenuous the upper boundary of a cloud is. Many analyses of CTH performance have focused on the column integrated cloud optical depth of a vertical profile. We suggest that CTH performance, and differences between CTH algorithms are more likely to be a function of the optical depth of (or the amount of water in) the uppermost portion of a cloud layer -- not the entire vertical profile.

Figure \ref{fig9} shows a comparison between C3DIR and ACHA using a CTH derived from the ACM-CAP product as ground truth. We compare the performance between C3DIR and ACHA as a function of the total mass of water in the column and the total amount of water in the uppermost 0.5 km of the cloud layer. The motivation for the latter is to perform the comparison for clouds where a significant portion of water in concentrated in the uppermost portion of a cloud, or for optically thick but geometrically thin cloud layers. Undoubtedly, this narrows the types of clouds represented in this analysis, but Figure \ref{fig9}.f shows this still represents a significant number of cloudy profiles.

Figures \ref{fig9}.a and \ref{fig9}.b show the MAE as a function of TWP in the profile and TWP in the uppermost 0.5 km for C3DIR and ACHA. Overall, C3DIR errors are all below roughly 2.5 km MAE with smaller errors with increasing TWP in the uppermost 0.5 km. A similar pattern is shown for ACHA but with higher errors throughout. Notably, errors vary much less as a function of TWP in the profile for both approaches. Figure \ref{fig9}.c shows that C3DIR matches ACM-CAP more closely throughout the comparison space, but particularly so for clouds with optically thin upper boundaries. The approaches become more similar with at least 10 \(\mathrm{g m^{-2}}\) of water in the uppermost 0.5 km of a cloud layer. Figures \ref{fig9}.d and \ref{fig9}.e show that ACHA typically has a slight low bias for clouds without thin upper boundaries, and C3DIR has a slight high bias for the same clouds.

\subsubsection{Hydrometeor-in-Profile Detection}

\begin{figure}[t]
    \centering
    \includegraphics[width=8cm]{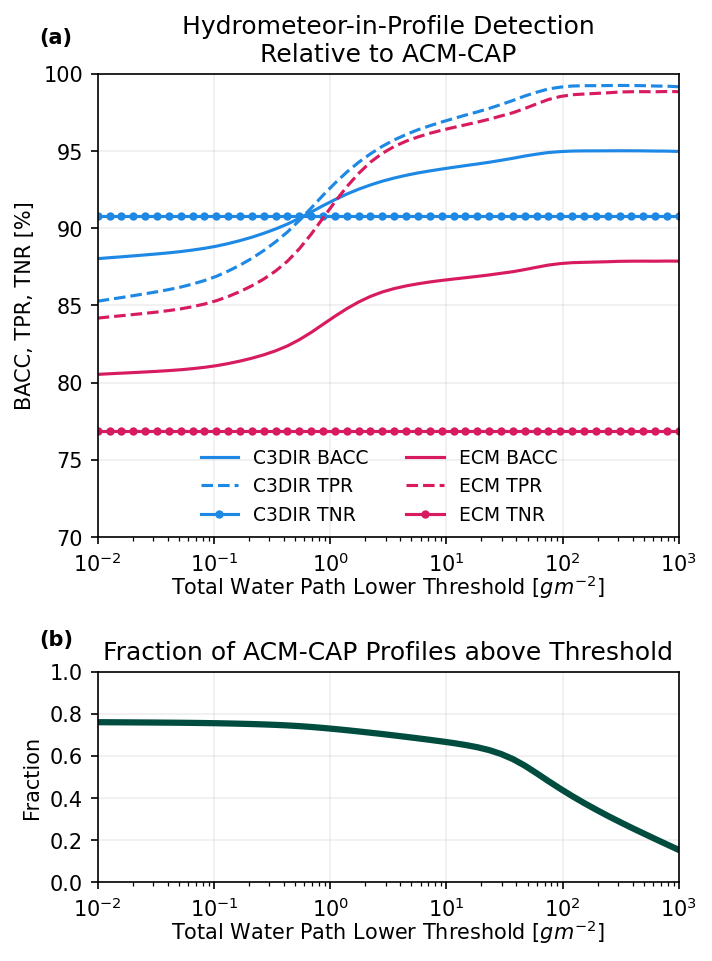}
    \caption{(a) An evaluation of C3DIR and ECM hydrometeor detection in vertical profiles as a function of a TWP lower threshold. At each point on the x-axis these metrics are evaluated on a subset of ACM-CAP profiles which include all clear-sky profiles and cloudy profiles with total water path exceeding the threshold. (b) shows the fraction of ACM-CAP profiles that exceed that threshold.}
    \label{fig10}
\end{figure}

Another more traditional cloud property estimated from passive imagers is the cloud mask. Intercomparisons between cloud masking algorithms are subject to many of the same caveats mentioned above. Cloud masking algorithms can be similarly tuned to different downstream applications which may have different acceptable amounts of cloud in clear-sky pixels. C3DIR is centered around a a very low water content threshold. We perform a limited comparison with the Enterprise Cloud Mask (ECM) which underpins the NOAA operational cloud masking product including only cloudy profiles above a minimum TWP. Figure \ref{fig10}.a illustrates the performance of a C3DIR cloud/hydrometeor mask compared to the ECM. BACC, TPR and TNR are calculated using the full set of cloud-free profiles from ACM-CAP and a subset of cloudy profiles where the TWP exceeds the value on the x-axis. Figure \ref{fig10} illustrates the ability of these methods to discriminate clouds with an increasing mass of water from completely clear-sky profiles in our collocated dataset.

As the TWP lower threshold increases, only cloudy profiles are excluded from the analysis keeping TNR constant throughout. C3DIR has a TNR of roughly 91.5\% while the ECM has a TNR of roughly 77\% indicating the C3DIR more reliably classifies cloud-free ACM-CAP profiles. However, it is possible that some of these cloud-free ACM-CAP profiles miss clouds very close to the surface that may have been caught by the ECM. Throughout the range of minimum TWP values, we note that C3DIR has a slightly higher TPR meaning it more reliably matches ACM-CAP in identifying cloudy profiles compared to the ECM. Overall, we find that that BACC is higher for C3DIR with the significant caveat that this result primarily driven by a much higher TNR that may not be representative of extremely low-level clouds.

\subsubsection{Column-Integrated Water Path}

\begin{figure}[t]
    \centering
    \includegraphics[width=16cm]{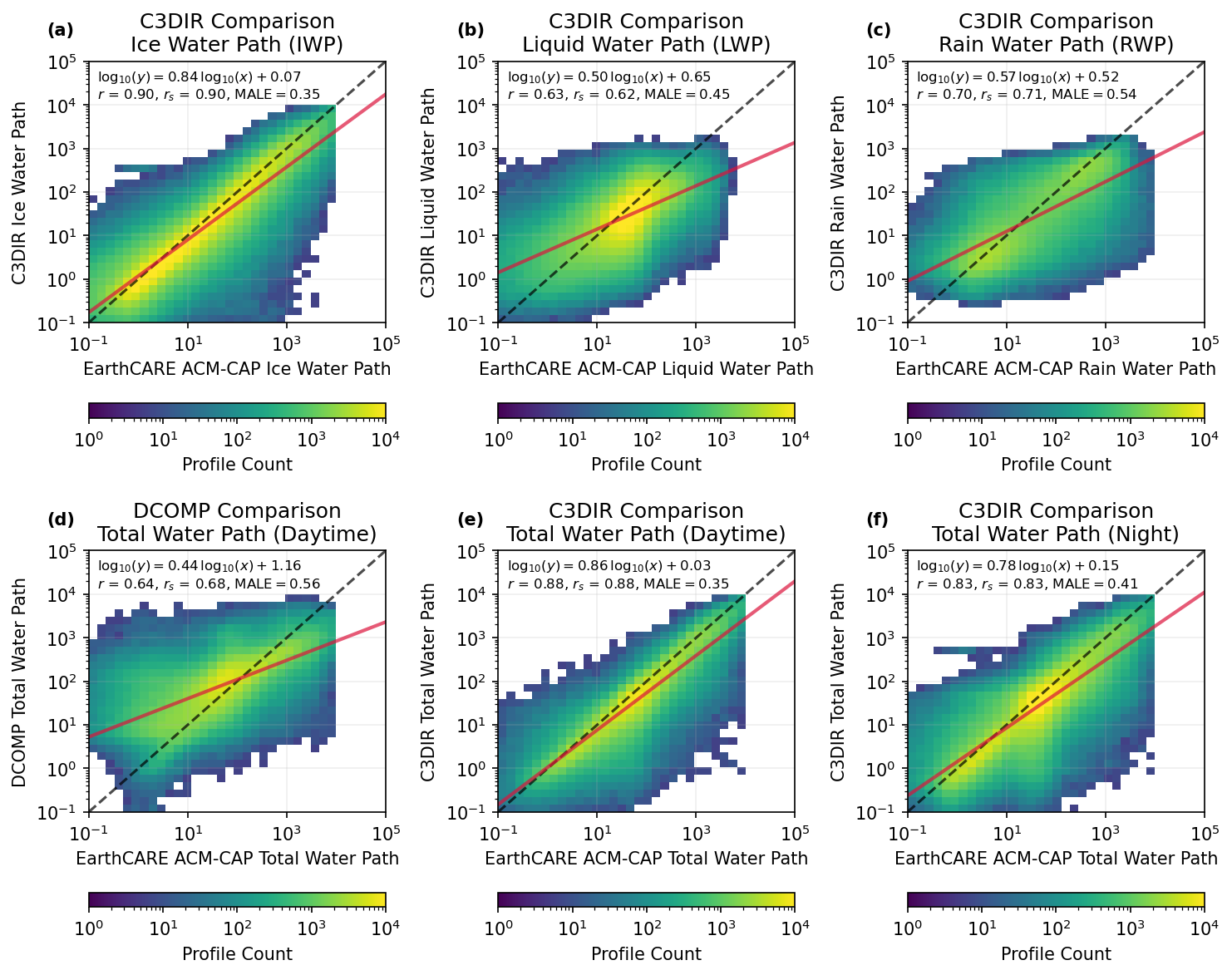}
    \caption{Evaluation of C3DIR and DCOMP water path with respect to ACM-CAP. (a)-(c) show the C3DIR ice, liquid, and rain water path. (d) and (e) show the DCOMP and C3DIR TWP during the daytime. (f) shows the C3DIR TWP during the night. The red line indicates a linear fit between log-scaled quantities. The text inset in each panel shows the Pearson and Spearman correlation coefficients, MALE and equation of the linear fit.}
    \label{fig11}
\end{figure}

More traditional approaches to estimating the total amount of cloud water in a column involve first estimating an optical depth and cloud-top effective radius, then calculating a water path based on those parameters \citep{han1994cwp,wood2006spatial,bennartz2007cwp}. This is in contrast to C3DIR's approach which estimates a conditional water content, masks out hydrometeor-free altitudes and integrates the water contents in the vertical direction.

In this analysis we compare C3DIR predictions and estimates of TWP derived from the Daytime Cloud Optical and Microphysical Properties algorithm (DCOMP; \citealp{walther2013atbd}) using optical depth and effective radius. Water Path from DCOMP is calculated as either IWP or LWP depending on the cloud-top phase using Eq. \ref{eq:lwp} and Eq. \ref{eq:iwp} as implemented in the current version of CLAVR-x, where $\tau$ is the cloud optical depth, $r_{e}$ is the effective radius, and $\rho_{liq}$ and $\rho_{ice}$ are the densities of liquid and ice water, respectively. Figures \ref{fig11}.a, b, and c show the water path for each species estimated by C3DIR. Note that vertical profiles often include a mix of water phases and these figures illustrate vertical integral of only those phases in C3DIR output. Ice water path appears to be very well characterized by C3DIR. Liquid and rain water path estimates have a moderate positive relationship with the EarthCARE ACM-CAP product, but still appear to less accurately match ACM-CAP compared to ice water path. Note that the water paths for all species have slightly higher correlation coefficients and lower MALE compared to the water contents from C3DIR. This implies that some of the challenge in estimating the water contents intuitively comes from predicting the altitude and vertical extent of clouds and precipitation.

\begin{equation}
    LWP = \frac{5\rho_{liq}\tau r_{e} }{9} 
\label{eq:lwp}
\end{equation}

\begin{equation}
    IWP = \frac{2\rho_{ice}\tau r_{e} }{3}
\label{eq:iwp}
\end{equation}

Figure \ref{fig11}.d and \ref{fig11}.e show DCOMP and C3DIR predictions for the same set of daytime EarthCARE collocations. Since DCOMP cannot distinguish the variation of water species throughout the vertical column we choose to compare the total water path performance between it and C3DIR. We filter daytime collocations to only profiles where the neighboring three vertical profiles on either side are contiguously cloudy to mitigate issues associated with cloud edges in DCOMP. Overall, we find that C3DIR has higher correlation coefficients and lower MALE compared to DCOMP with respect to ACM-CAP. Notably, DCOMP has a large number of significant overestimates for thin clouds identified by ACM-CAP. One possibility is that these represent multi-layer scenes where thin upper-level cirrus overlies very low-level cumulus undetected by ACM-CAP. To check this possibility, we re-run these comparisons after removing clouds with CLAVR-x CBH below 1 km and 2 km (see Table \ref{tab:dcomp_base}). Overall, we find that performance increases from both approaches, but C3DIR still outperforms DCOMP for the remaining clouds. We expect it is likely that undetected boundary layer clouds contribute to some of the DCOMP error in this analysis, but removing such clouds identified in the CBH estimates does not account for the entire difference between C3DIR and DCOMP.

Figure \ref{fig11}.f illustrates the performance of TWP from C3DIR at night. There is only a small difference in the performance metrics when evaluating TWP from C3DIR during the daytime. This is possibly a sign that C3DIR is not making use of the information in the VIS and NIR channels that possess stronger physical sensitivity to optical depth and particle size. Alternatively, recent work has shown that spatial context can significantly improve the estimation of cloud and precipitation\citep{jeggle2023spatial,hilburn2023spatial}, and can make up for the lack of sensitivity to optically thick clouds in IR channels \citep{white2025unetcomp}.

\begin{table}[]
\centering
\begin{tabular}{lllllll}
                      &      & DCOMP  &      &      & C3DIR  &      \\ \cline{2-7} 
                      & $r$    & $r_{s}$ & MALE & $r$    & $r_{s}$ & MALE \\ \hline
All CBH               & 0.64 & 0.68   & 0.56 & 0.88 & 0.88   & 0.35 \\
CBH \textgreater 1 km & 0.70 & 0.73   & 0.47 & 0.88 & 0.89   & 0.34 \\
CBH \textgreater 2 km & 0.72 & 0.76   & 0.45 & 0.89 & 0.90   & 0.33 \\ \hline
\end{tabular}
\vspace{0.5em}
\caption{Comparison of C3DIR and DCOMP daytime TWP to EarthCARE ACM-CAP considering only clouds with increasingly elevated cloud base identified from the NOAA operational CBH algorithm. $r$ and $r_{s}$ are the Pearson and Spearman correlation coefficients, and MALE is the mean absolute log error. }
\label{tab:dcomp_base}
\end{table}

\subsubsection{Detection of Multiple Layers}

\begin{figure}[t]
    \centering
    \includegraphics[width=16cm]{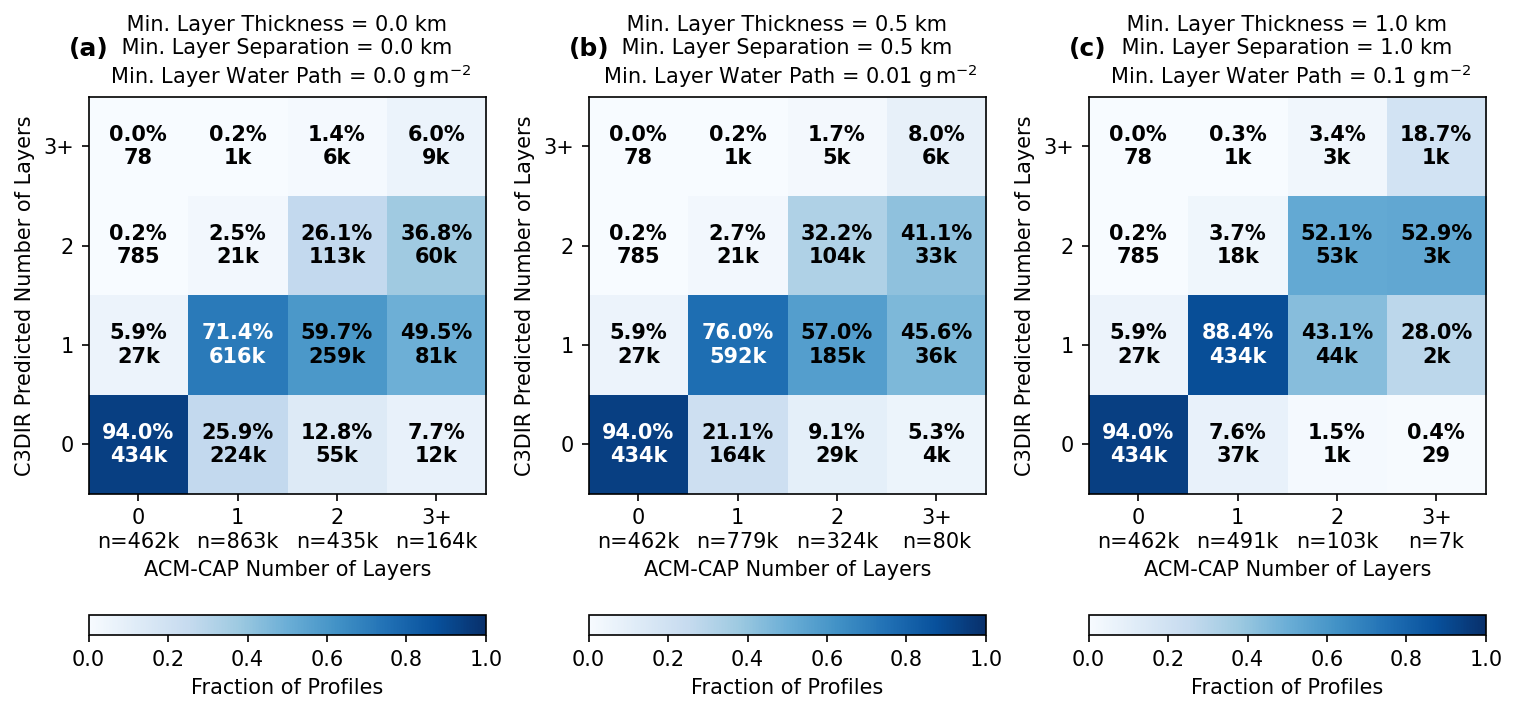}
    \caption{Confusion matrices of the predicted number of layers from C3DIR compared to ACM-CAP. (a) sets no requirements on the minimum geometric thickness, minimum separation between layers or the minimum amount of water in a cloud layers. (b) and (c) set increasingly strict requirements listed above each panel. Note that these requirements must apply to all layers in a given profile in order to be included here. Percentages are a function of the number of samples in a a given column with the totals listed along the x-axis.}
    \label{fig12}
\end{figure}

Given the objectives of C3DIR in rendering 3-D volumes of clouds from 2-D imagery, it is helpful to quantify the ability of the model to detect multiple distinct layers in the vertical dimension. There are several different considerations when performing such an analysis depending on the exact downstream application or use case. Here we consider multiple requirements regarding the minimum geometric thickness, minimum vertical separation between layers, and the minimum water path in a cloud. Rather than combining individual layers in a profile that don't meet these requirements, we choose more simply to perform our analysis on a subset of vertical profiles where all layers meet these requirements. 

Figure \ref{fig12} shows the confusion matrices under increasing strict requirements on the multi-layered profiles in our testing dataset. Under the most permissive requirements, it is clear that C3DIR has a strong tendency to underestimate the number of cloud layers. This is unsurprising given the number of geometrically thin layers with small horizontal scale present in Figure \ref{fig5}.g, for example, that C3DIR significantly smooths over. Intuitively, as the minimum layer thickness, minimum clear-sky separation between layers, and minimum water path in a given layer increases, so does the C3DIR performance in identifying the presence of multiple layers. However, we note that the statistics shown in Figure \ref{fig12}.c come from an increasingly small number of vertical profiles due to the need for all layers in a profile to satisfy those conditions. 

\subsection{Ground-based intercomparisons}
\label{sec:ground_based}

The analysis so far has only evaluated C3DIR against the EarthCARE ACM-CAP product it was trained to match using several NOAA operational products as baselines where applicable. Next, we perform similar comparisons using ground-based remote sensing instrumentation available at the SGP-ARM site. This comparison is intended to assess whether C3DIR has overfit to any biases present in the EarthCARE ACM-CAP product that may compromise our assessments above with respect to the NOAA operational products. 

\begin{figure}[t]
    \centering
    \includegraphics[width=16cm]{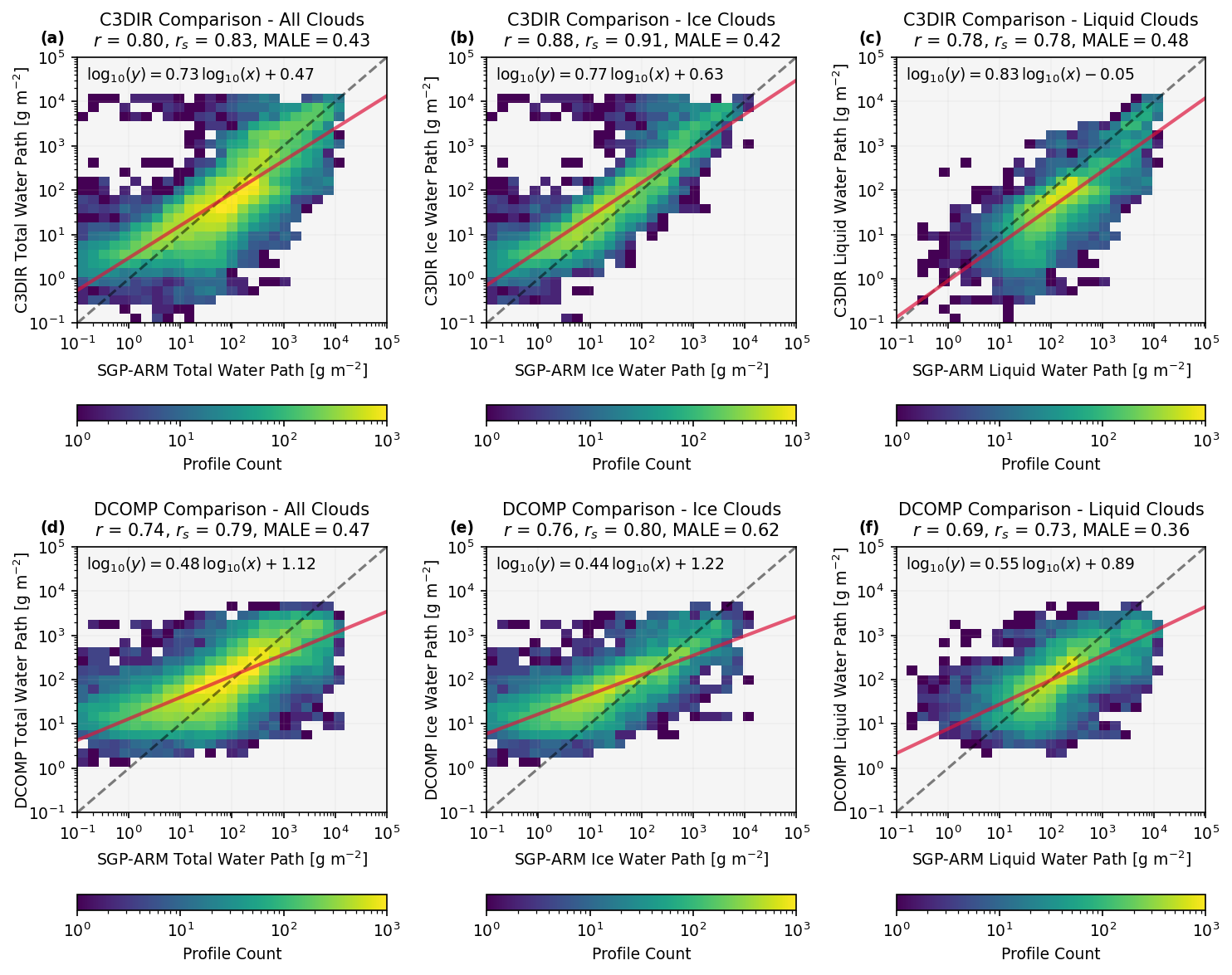}
    \caption{Evaluation of C3DIR and DCOMP water path with respect to the SGP-ARM site estimates from the Microbase product. (a)-(c) show the C3DIR TWP, IWP and the sum total of LWP and RWP. (d)-(f) show the same quantities for DCOMP. The red line indicates a linear fit between log-scaled quantities. Text inset in each panel shows the Pearson and Spearman correlation coefficients, MALE and equation of the linear fit. Note that for ice clouds (b,c) and liquid clouds (c,f), we restrict our analysis to profiles in which greater than 90\% of the TWP belongs to each phase.}
    \label{fig13}
\end{figure}

Figure \ref{fig13} compares both C3DIR and DCOMP to vertically-integrated water path from the SGP-ARM profiles. Figures \ref{fig13}.a and \ref{fig13}.d specifically illustrate comparisons among both liquid and ice clouds. We note that both the MALE for C3DIR and DCOMP are roughly similar and the correlation coefficients are slightly higher for C3DIR.  When restricting our analysis to only profiles where greater than 90\% of the TWP is attributed to ice-phase water, C3DIR more strongly outperforms DCOMP. However, when restricting our analysis to profiles where greater than 90\% of the TWP is attribute to liquid-phase water (including precipitation), we find that C3DIR has higher correlation coefficients, but larger MALE compared to DCOMP. These larger errors can be attributed to a somewhat consistent low bias seen in the C3DIR estimates of liquid phase water content with respect to the SGP-ARM profiles. From this analysis alone, it is unclear whether this is a bias present in the SGP-ARM profiles, a feature of the EarthCARE ACM-CAP product inherited by C3DIR, or generalization error from C3DIR. Such an assessment is made difficult by the fact that the SGP-ARM comparisons occur over just a single location, while the ACM-CAP comparisons are performed over a globally-distributed set of collocations. Future work in characterizing the performance of these algorithms from in-situ observations or other instrument platforms would be beneficial towards resolving these differences. 

\begin{figure}[t]
    \centering
    \includegraphics[width=16cm]{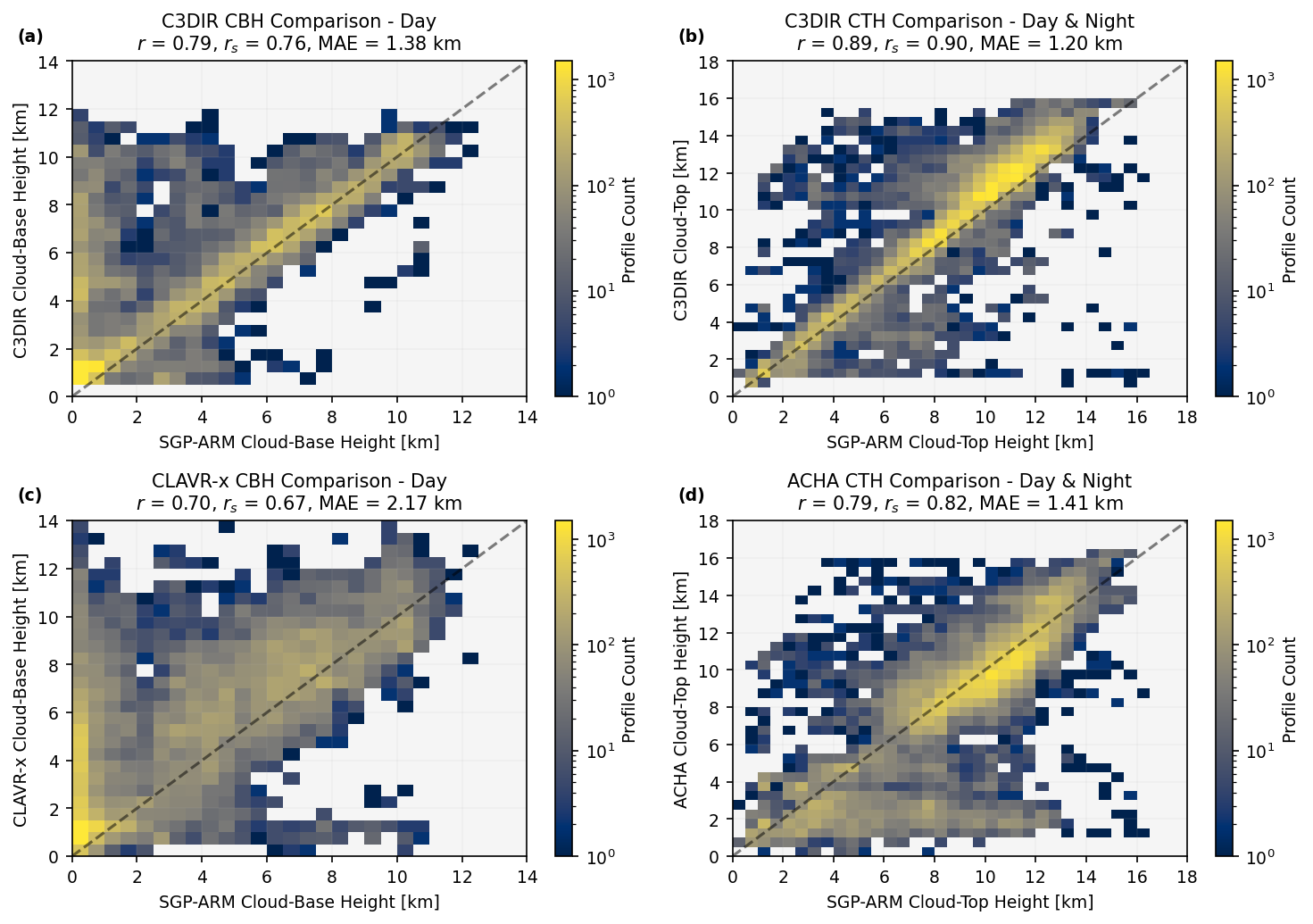}
    \caption{Evaluation of CTH and CBH from C3DIR and the NOAA operational algorithms with respect to those from the Microbase product at the SGP-ARM site. (a) and (c) show C3DIR and CLAVR-x CBH during the daytime. (b) and (d) show C3DIR and ACHA CTH during all times of day.}
    \label{fig14}
\end{figure}

Comparisons of CTH and CBH using the SGP-ARM site products are shown in Figure \ref{fig14}. For the purposes of this evaluation we apply a similar threshold of $1\times10^{-5}$ \(\mathrm{g m^{-3}}\) to delineate clear-sky portions of vertical profiles from hydrometeors. We treat cloud-base height as the lowest altitude with hydrometeors present with no special handling of multi-layered scenes. Comparisons of CBH are limited to daytime only since the NOAA operational CBH product relies on accurate estimates of cloud water path that are currently of limited quality at night. C3DIR CBH values are available at all times of day, but we similarly limit this evaluation to daytime profiles to facilitate comparison. CBH estimates from C3DIR overall align more closely with SGP-ARM with lower MAE and higher correlation coefficient than those from the NOAA operational products. Similar to the EarthCARE ACM-CAP comparison, the largest errors occur where SGP-ARM instrumentation places cloud-base just above the surface, and the satellite algorithms place it much higher in the vertical profile. Aside from those cases, Figures \ref{fig14}.a and \ref{fig14}.c show that C3DIR CBH clusters more tightly around the 1:1 line compared to the NOAA operational product.

C3DIR CTH agrees relatively well with the SGP-ARM site. We note that C3DIR CTH is consistently biased slightly higher than the SGP-ARM products, with a few cases where C3DIR appears to falsely identify upper-level clouds. ACHA CTH are typically less tightly clustered around the 1:1 line. ACHA also has a collection of profiles where ACHA places CTH at 1-3 km where the SGP-ARM instrumentation identifies a mid- or upper-level cloud. We expect that these are multi-layered profiles with optically thin upper layers that are not identified by ACHA. Overall, C3DIR upper and lower cloud boundaries are in closer agreement with the SGP-ARM site compared to the NOAA operational products.

\begin{figure}[t]
    \centering
    \includegraphics[width=16cm]{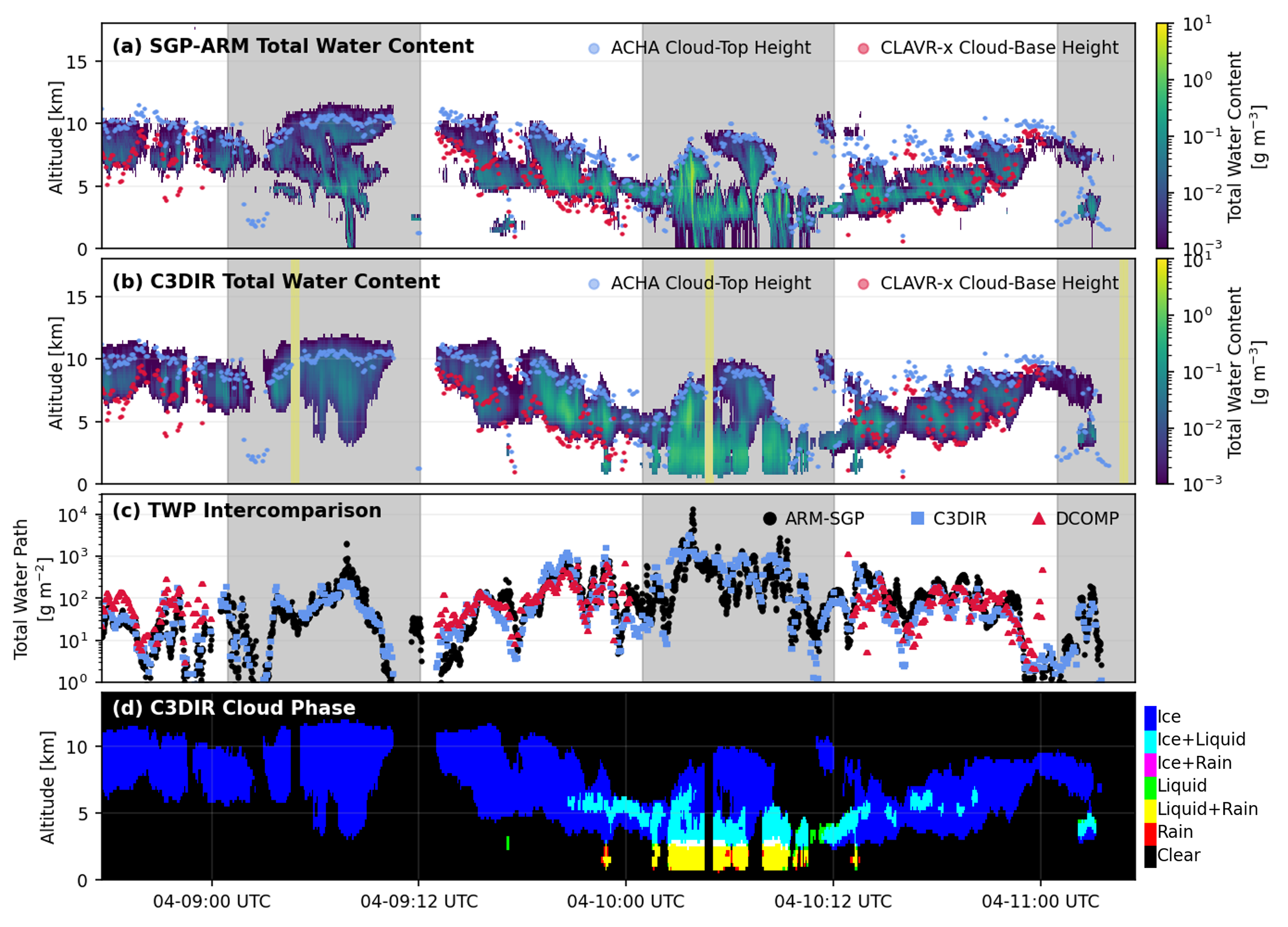}
    \caption{Comparison of vertical cross sections from the SGP-ARM site MICROBASE products. (a) shows the TWC from the SGP-ARM Microbase products. (b) shows TWC from C3DIR. (c) shows the vertically integrated TWP from the SGP-ARM site, C3DIR and DCOMP. (d) shows combination of ice, liquid and rain masks from C3DIR. The pale yellow background in (b) represents times where any of the ABI channels were not available at this location, thus no prediction from C3DIR is made.}
    \label{fig15}
\end{figure}

Next, we turn towards a more qualitative comparison between the vertical profiles from the SGP-ARM site and C3DIR. Figure \ref{fig15}.a and \ref{fig15}.b show profiles of TWC with the NOAA operational CTH and CBH overlaid on top for comparison. Overall, both the NOAA operational products and C3DIR show good agreement with the SGP-ARM profiles in terms of cloud boundaries with a few exceptions. During the night at around 6 UTC April 9th, C3DIR fails to characterize the multi-layered profiles with relatively small cloud-free gaps between them. C3DIR instead smooths these clouds out in the vertical direction. The NOAA operational products do not estimate a cloud base at these times due to the lack of cloud water path information at night from the operational products. 

During the daytime between 12 UTC April 9th, 2024 to 0 UTC April 10th, C3DIR and the SGP-ARM products agree fairly well in terms of CTH, CBH and water content. A low-level cloud liquid cloud at around 3 km altitude is missed by C3DIR beneath the upper level ice clouds at around 17 UTC on April 9th. At this same time, the NOAA operational cloud-top and base seem characterize a middle ground between the upper-level cloud at and lower-level cloud, and correctly identify the boundaries at the lower level in a small number of profiles. The SGP-ARM product identifies precipitation reaching the surface during the night between 0 UTC April 10, and 12 UTC April 10. C3DIR correctly identifies the cloud-top location, and water content in the uppermost portions of these clouds, and the presence of precipitation. However, it severely overestimates the water content of the precipitation relative to the SGP-ARM products and does not extend that precipitation down to the surface. The presence of precipitation seems to complicate the characterization of C3DIR TWC throughout most of the column below the uppermost portion of the cloud.

After 12 UTC April 10, C3DIR cloud boundaries and water content match the SGP-ARM site fairly well. The NOAA operational products are more inconsistent during this time period. Across the two days shown in Figure \ref{fig15}, the NOAA CBH appears to be more inconsistent at dawn and dusk. We expect this is due to higher solar zenith angles that complicate the estimation of the cloud water path used in the CBH algorithm. 

Throughout the two day period shown we note that when cloud-tops in the SGP-ARM profiles have relatively small water content, ACHA places the cloud-top somewhere within the cloud. This is consistent with the analysis shown in Figure \ref{fig9}. We also note that C3DIR seems to reliably characterize the low values of water content near cloud top. We see this as a promising sign that cloud-top definitions could be tuned to downstream applications based on the estimated water content thresholds from C3DIR. However, we caution that this aspect of C3DIR output requires further evaluation. 

\section{Discussion}
\label{sec:discussion}

Overall, C3DIR appears to estimate cloud properties in good agreement with the EarthCARE ACM-CAP product, but with some significant caveats. Hydrometeor detection among individual voxels performs well with respect to ACM-CAP, especially given the limitations of passive satellite measurements. At the voxel level, C3DIR shows a tendency to under predict the presence of hydrometeors evidenced by a slightly higher TNR compared to TPR across all altitudes for each species of water (Figure \ref{fig6}). This a consequence of the extremely low water content threshold we use to define the presence or absence of hydrometeors in a voxel. Intuitively, we see that hydrometeor detection performance is strongly a function of the TWC in a given voxel. However, this association is much weaker for liquid clouds (Figure \ref{fig6}.g) which are difficult in part because of their finer horizontal spatial scale and small geometric thickness.

The voxel-level water content from C3DIR agrees well with ACM-CAP for total and ice water content, but less so for liquid and rain water content. We expect that the potential explanations for weaker liquid voxel detection performance also apply to the estimation of liquid and rain water content. In addition, liquid and rain water content are among the more challenging quantities to characterize in the ACM-CAP retrieval. Liquid cloud retrievals are exceedingly difficult in multilayer and mixed-phase profiles due to lidar attenuation, radar limitations, and uncertainty in phase classification which all weaken constraints on the estimated state \citep{mason2023acmcap, mason2024comparison}. We expect that the performance and interpretation of C3DIR's liquid and rain water content output to improve as validation work for the EarthCARE ACM-CAP products continues. 

At the profile-level we find that the C3DIR can reliably estimate CTH and outperform the current NOAA operational product (ACHA) when using our coarsened ACM-CAP dataset as ground truth. Much of the difference in performance comes from clouds with tenuous upper boundaries and could be attributed to differing definitions of what constitutes the top of a cloud. When controlling for clouds with relatively large amounts of water in the uppermost 0.5 km of the cloud, we find there is still consistent improvement in C3DIR, but the difference in performance is much smaller. Similar conclusions can be drawn from our profile-level hydrometeor detection comparison (Figure \ref{fig10}). Overall, we find that C3DIR more reliably discriminates clear-sky from cloudy profiles. However, there is a strong possibility that C3DIR and the ECM are centered around fundamentally different definitions of what constitutes a cloudy imager pixel due to the large differences in TNR and the very low water content threshold we use to define cloudy from clear-sky in C3DIR. Despite this potential difference, we find that C3DIR consistently has a higher TPR for clouds with high TWP where we would otherwise expect cloud masks to be in very strong agreement regardless of their respective cloud definitions.

Comparisons with a ground-based dataset at the SGP-ARM site confirm that C3DIR can reliably estimate boundaries of clouds (e.g, CTH and CBH) and seemingly outperforms the NOAA operational products. Ice water path from C3DIR is in particularly good agreement with the MICROBASE products. However we note that for liquid water path, C3DIR has a substantial low bias but a stronger linear relationship with SGP-ARM compared to DCOMP. The vertical cross sections from C3DIR at the SGP-ARM site in Figure \ref{fig15} appear to be of slightly lower quality compared to cross sections with EarthCARE in Figure \ref{fig4} and \ref{fig5}. The majority of EarthCARE collocations occur over open ocean, so it is possible that C3DIR performance differs over land. Future evaluations and updates to C3DIR may benefit from characterizing performance over different geographic regions and surface types once a longer EarthCARE record is available. Our EarthCARE testing set covers a limited time period due to the size of the training set needed to train the model, so the analysis presented here focuses on global aggregate statistics.

C3DIR has shown improvement over several of the NOAA operational products in this work, but we do not see that as the primary benefit of using this model. That factor alone does not necessarily justify the cost and complexity that accompanies C3DIR. Rather, we see the primary benefits of C3DIR to be what it offers beyond what is available in the current suite of operational products. C3DIR has the ability to characterize vertically-resolved cloud properties rather than more simple profile-integrated water content, or cloud-top properties. Similarly, C3DIR can also estimate multiple distinct layers and the associated water contents may allow for more flexible use of the output for downstream applications. For example, one might filter out upper-level thin cirrus according to an application-specific cloud definition focused on higher water content thresholds. The 3-D output field can be optionally used to construct vertical profiles normal to the surface by resampling voxels from multiple imager pixels. This is demonstrated in our comparisons to EarthCARE. This may also facilitate comparisons between multiple imagers and may reduce the need for simultaneous nadir overpasses (SNOs) or otherwise ray-matched collocations. The more precise 3-D collocation methodology that allows for this also facilitates training and evaluating model output at higher viewing angles closer to the edge of the full-disk or scan.

The analysis here provides globally-aggregated statistics and the general behavior of C3DIR with respect to the ACM-CAP product. There are several aspects of the model design in C3DIR that would benefit from future investigation. We note again that C3DIR has roughly similar performance in estimating TWP between day and night in Figure \ref{fig12}. There is a possibility that C3DIR is not relying on VIS or NIR information that would be expected to benefit the characterization of optically thick clouds. This could potentially be remedied by including solar or viewing geometry in the inputs to account for the overall brightness of the scene or cloud shadowing and other 3-D radiative effects. The satellite-specific stem gives the model the capability of learning the unique spectral response of individual sensors. However, it is not clear how large this component of the model needs to be, or to what degree channels unique to each satellite are used. During the development of C3DIR, we did not notice substantial differences in performance between sensors, but this is complicated by the different geographic areas viewed and requires more investigation.

There are a number of operations-related considerations as well. As with many other supervised AI/ML-based models for 3-D cloud properties, C3DIR requires collocations with active sensor vertical profiles. In cases where a newly launched imaging instrument needs to be added to C3DIR, there are a few possibilities. The ideal case is that the new sensor has a large set of collocations with EarthCARE. In such a case, C3DIR could be entirely retrained for all imagers. If the record of EarthCARE collocations with the new sensor is small, it is possible that only a satellite-specific stem for the new sensor would need to be trained with the shared encoder and decoder weights frozen. Alternatively, the new sensor could be trained to emulate performance of an existing sensor if there is geographic overlap between the two. If there are no collocations with active sensor (e.g., in the case of developing algorithms for imagers that have not yet launched), then the task is significantly more difficult. We expect that C3DIR and other efforts will benefit from future research targeting the use of unsupervised or semi-supervised learning to support operational use of newly launched records without long records of collocations with active vertical profiling instruments.

Another consideration is the computational cost of running C3DIR. As mentioned previously, it takes a total of 220 GPU-hours to train the version of C3DIR presented here using four NVIDIA RTX A6000s in parallel. In some ways, this is a one-time cost for a given version model. However we expect ongoing development and the addition of new sensors to require multiple re-training runs. After it has been trained, the presented version of C3DIR can produce a 3-D volume for the entire ABI full disk in less than two minutes. While, this may be fast enough to support operational use, we emphasize that this requires a GPU, which is not a requirement for any of the operational products discussed here.

With all of the above taken together, these results suggest that C3DIR may provide a path toward retrieving vertically resolved cloud structures from passive satellite imagers. Several uncertainties remain regarding liquid and rain water content, but C3DIR captures much of the large-scale behavior of the ACM-CAP product and compares favorably with selected operational products. The contribution of C3DIR is its ability to characterize cloud vertical structure, multiple layers, and vertically resolved water content in a form that we expect can be adapted to application-specific requirements. Continued evaluation with longer records and more extensively-validated products from EarthCARE, additional intercomparisons with other instrument platforms, and evaluations targeting specific components of C3DIR will be necessary to determine operational utility and robustness of the presented model.

\section{Conclusions}
\label{sec:conclusions}

In this study, we develop the Cloud 3-Dimensional Imager Retrieval (C3DIR), a deep learning model that estimates 3-D cloud properties from conventional passive imaging instruments, using training data from the recently launched EarthCARE satellite. We illustrate a voxel-to-voxel collocation procedure in which C3DIR is trained to estimate cloud properties along a given passive imager's line-of-sight through the atmosphere. Unlike traditional profile-to-pixel matching methods, this approach provides a more precise collocation of EarthCARE observations and 2D pixel-based passive sensor data by matching portions of the active vertical profile directly with the output space of a 3-D cloud property retrieval.

C3DIR is capable of characterizing cloud vertical structure from multiple passive imager measurements in broad agreement with the EarthCARE ACM-CAP product. Qualitative comparisons with ACM-CAP cross-sections show C3DIR accurately capturing layered clouds including distinct liquid-topped mixed phase clouds beneath upper-level ice and thin cirrus overlying deep convection. The model performs well in voxel-level hydrometeor detection and shows especially strong agreement for the detection of ice and the estimation of ice water content. Profile-level comparisons indicate reliable estimates of cloud-top height and base height, cloud detection, and column-integrated water path which compare favorably to current operational products. However, liquid and rain water content estimates from C3DIR remain more uncertain, reflecting both limitations of passive VIS/NIR/IR remote sensing and the difficulty of retrieving some of these quantities from active sensor products. Comparisons with operational products highlight several potential areas where C3DIR improves upon them, but these differences partly reflect different effective definitions of cloud occurrence rather than algorithm performance alone. 

The primary value of C3DIR is in its ability to provide a flexible 3-D representation of cloud structure including the identification of multiple cloud layers and vertically resolved water content for any of the satellite imagers included in its training data. We show that one can resample voxels from multiple imager pixels in C3DIR's output to construct vertical cross-sections of the atmosphere aligned with the near-nadir views from EarthCARE instruments. This capability reduces the dependence on single-pixel, single-column interpretations of passive imager estimates. We expect that this model will benefit from continued evaluation and training over longer EarthCARE records and more focused comparisons over different surface types, geographic regions, viewing geometry, and cloud regimes. Overall, these results demonstrate the potential for C3DIR to offer broader utility for downstream applications. 

\section*{Acknowledgments}

This work was supported by the Office of Naval Research under award N0001424C2214, and the NOAA GOES and JPSS Program offices under Cooperative Agreement NA24OARX432C0007. 

We acknowledge the European Space Agency (ESA), the Japan Aerospace Exploration Agency (JAXA), and the EarthCARE science team for making their products publicly available. This study uses observations and products from the Southern Great Plains atmospheric observatory, provided by the U.S. Department of Energy Atmospheric Radiation Measurement (ARM) user facility. This work used observations from the GOES, Himawari, and Meteosat geostationary satellite programs. The authors gratefully acknowledge NOAA for GOES ABI data, JMA for Himawari AHI data, and EUMETSAT for Meteosat SEVIRI data.

We would also like to thank Cassidy M. Johnson (CIRA) and Steven M. Miller (CIRA) for enlightening discussions on the use of active sensor observations, and potential improvements and applications of C3DIR.

All data used in this work are publicly available at no cost from their respective agencies. The algorithms behind the NOAA operational product are run using publicly available CLAVR-x software \\ (\url{https://cimss.ssec.wisc.edu/clavrx/documentation/}).


\newpage
\bibliographystyle{plainnat_modified}
\bibliography{references}

@article{bony2015clouds,
    title={{Clouds, circulation and climate sensitivity}},
    author={Bony, Sandrine and Stevens, Bjorn and Frierson, Dargan MW and Jakob, Christian and Kageyama, Masa and Pincus, Robert and Shepherd, Theodore G and Sherwood, Steven C and Siebesma, A Pier and Sobel, Adam H and others},
    journal={Nature Geoscience},
    volume={8},
    number={4},
    pages={261--268},
    year={2015},
    publisher={Nature Publishing Group UK London},
    url={https://doi.org/10.1038/ngeo2398}
}

@inbook{ipccar6,
    place={Cambridge},
    title={{The Earth’s Energy Budget, Climate Feedbacks and Climate Sensitivity}},
    booktitle={Climate Change 2021 – The Physical Science Basis: Working Group I Contribution to the Sixth Assessment Report of the Intergovernmental Panel on Climate Change},
    publisher={Cambridge University Press},
    author={{Intergovernmental Panel on Climate Change (IPCC)}},
    year={2023},
    pages={923–1054},
    url={https://doi.org/10.1017/9781009157896.009},
}

@article {stephens2005review,
    author={Stephens, Graeme L},
    title = {{Cloud Feedbacks in the Climate System: A Critical Review}},
    journal = "Journal of Climate",
    year = "2005",
    publisher = "American Meteorological Society",
    address = "Boston MA, USA",
    volume = "18",
    number = "2",
    url = {https://doi.org/10.1175/JCLI-3243.1},
    pages= "237 - 273",
}

@article{oreopoulos2017cvs,
    author = {Oreopoulos, Lazaros and Cho, Nayeong and Lee, Dongmin},
    title = {{New insights about cloud vertical structure from CloudSat and CALIPSO observations}},
    journal = {Journal of Geophysical Research: Atmospheres},
    volume = {122},
    number = {17},
    pages = {9280-9300},
    url = {https://doi.org/10.1002/2017JD026629},
    year = {2017}
}

@article{cahalan1994fractional,
    author={Cahalan, Robert F and Ridgway, William and Wiscombe, Warren J and Bell, Thomas L and Snider, Jack B},
    title = {{The Albedo of Fractal Stratocumulus Clouds}},
    journal = "Journal of Atmospheric Sciences",
    year = "1994",
    publisher = "American Meteorological Society",
    address = "Boston MA, USA",
    volume = "51",
    number = "16",
    url = {https://doi.org/10.1175/1520-0469(1994)051<2434:TAOFSC>2.0.CO;2},
    pages= "2434 - 2455",
}

@article{Zhang2011horizontal,
    author = {Zhang, Zhibo and Platnick, Steven},
    title = {{An assessment of differences between cloud effective particle radius retrievals for marine water clouds from three MODIS spectral bands}},
    journal = {Journal of Geophysical Research: Atmospheres},
    volume = {116},
    number = {D20},
    pages = {},
    url = {https://doi.org/10.1029/2011JD016216},
    year = {2011}
}

@article{Barker1999geometry,
    author = {Barker, Howard W. and Stephens, Graeme L. and Fu, Qiang},
    title = {{The sensitivity of domain-averaged solar fluxes to assumptions about cloud geometry}},
    journal = {Quarterly Journal of the Royal Meteorological Society},
    volume = {125},
    number = {558},
    pages = {2127-2152},
    url = {https://doi.org/10.1002/qj.49712555810},
    year = {1999}
}

@article{barker2008cloudsat,
    author = {Barker, Howard W.},
    title = {{Representing cloud overlap with an effective decorrelation length: An assessment using CloudSat and CALIPSO data}},
    journal = {Journal of Geophysical Research: Atmospheres},
    volume = {113},
    number = {D24},
    pages = {},
    url = {https://doi.org/10.1029/2008JD010391},
    year = {2008}
}

@article{cao2018icing,
    author={Cao, Yihua and Tan, Wenyuan and Wu, Zhenlong},
    title = {{Aircraft icing: An ongoing threat to aviation safety}},
    journal = {Aerospace Science and Technology},
    volume = {75},
    pages = {353-385},
    year = {2018},
    issn = {1270-9638},
    url = {https://doi.org/10.1016/j.ast.2017.12.028},
}

@article{miller2018solar,
    author={Miller, Steven D and Rogers, Matthew A and Haynes, John M and Sengupta, Manajit and Heidinger, Andrew K},
    title = {{Short-term solar irradiance forecasting via satellite/model coupling}},
    journal = {Solar Energy},
    volume = {168},
    pages = {102-117},
    year = {2018},
    note = {Advances in Solar Resource Assessment and Forecasting},
    issn = {0038-092X},
    url = {https://doi.org/10.1016/j.solener.2017.11.049},
}

@article{paletta2023solar,
    author={Paletta, Quentin and Terr{\'e}n-Serrano, Guillermo and Nie, Yuhao and Li, Binghui and Bieker, Jacob and Zhang, Wenqi and Dubus, Laurent and Dev, Soumyabrata and Feng, Cong},
    title = {{Advances in solar forecasting: Computer vision with deep learning}},
    journal = {Advances in Applied Energy},
    volume = {11},
    pages = {100150},
    year = {2023},
    issn = {2666-7924},
    url = {https://doi.org/10.1016/j.adapen.2023.100150},
}

@article{hoffman2024dmw,
    author = {Hoffman, Ross N. and Liu, Hui and Lukens, Katherine E. and Garrett, Kevin and Ide, Kayo},
    title = {{Assimilating atmospheric motion vector winds using a feature track correction observation operator}},
    journal = {Quarterly Journal of the Royal Meteorological Society},
    volume = {150},
    number = {765},
    pages = {5074-5093},
    url = {https://doi.org/10.1002/qj.4857},
    year = {2024}
}

@article{velden2009dmv,
    author={Velden, Christopher S and Bedka, Kristopher M},
    title = {{Identifying the Uncertainty in Determining Satellite-Derived Atmospheric Motion Vector Height Attribution}},
    journal = "Journal of Applied Meteorology and Climatology",
    year = "2009",
    publisher = "American Meteorological Society",
    address = "Boston MA, USA",
    volume = "48",
    number = "3",
    url = {https://doi.org/10.1175/2008JAMC1957.1},
    pages= "450 - 463",
}

@article{heidinger2012nbcm,
    author={Heidinger, Andrew K and Evan, Amato T and Foster, Michael J and Walther, Andi},
    title = {{A Naive Bayesian Cloud-Detection Scheme Derived from CALIPSO and Applied within PATMOS-x}},
    journal = "Journal of Applied Meteorology and Climatology",
    year = "2012",
    publisher = "American Meteorological Society",
    address = "Boston MA, USA",
    volume = "51",
    number = "6",
    url = {https://doi.org/10.1175/JAMC-D-11-02.1},
    pages="1129 - 1144",
}

@article{mecikalski2007satellites,
    author={Mecikalski, John R and Feltz, Wayne F and Murray, John J and Johnson, David B and Bedka, Kristopher M and Bedka, Sarah T and Wimmers, Anthony J and Pavolonis, Michael and Berendes, Todd A and Haggerty, Julie and others},
    title = {{Aviation Applications for Satellite-Based Observations of Cloud Properties, Convection Initiation, In-Flight Icing, Turbulence, and Volcanic Ash}},
    journal = "Bulletin of the American Meteorological Society",
    year = "2007",
    publisher = "American Meteorological Society",
    address = "Boston MA, USA",
    volume = "88",
    number = "10",
    url = {https://doi.org/10.1175/BAMS-88-10-1589},
    pages="1589 - 1607",
}

@article{frey2020mvcm,
    AUTHOR = {Frey, Richard A. and Ackerman, Steven A. and Holz, Robert E. and Dutcher, Steven and Griffith, Zach},
    TITLE = {{The Continuity MODIS-VIIRS Cloud Mask}},
    JOURNAL = {Remote Sensing},
    VOLUME = {12},
    YEAR = {2020},
    NUMBER = {20},
    ARTICLE-NUMBER = {3334},
    ISSN = {2072-4292},
    url = {https://doi.org/10.3390/rs12203334}
}

@article{hunderbein2023mask,
    AUTHOR = {H\"unerbein, A. and Bley, S. and Horn, S. and Deneke, H. and Walther, A.},
    TITLE = {{Cloud mask algorithm from the EarthCARE Multi-Spectral Imager: the M-CM products}},
    JOURNAL = {Atmospheric Measurement Techniques},
    VOLUME = {16},
    YEAR = {2023},
    NUMBER = {11},
    PAGES = {2821-2836},
    url = {https://doi.org/10.5194/amt-16-2821-2023}
}

@article{bulgin2018errors,
    AUTHOR = {Bulgin, Claire E. and Merchant, Christopher J. and Ghent, Darren and Klüser, Lars and Popp, Thomas and Poulsen, Caroline and Sogacheva, Larisa},
    TITLE = {{Quantifying Uncertainty in Satellite-Retrieved Land Surface Temperature from Cloud Detection Errors}},
    JOURNAL = {Remote Sensing},
    VOLUME = {10},
    YEAR = {2018},
    NUMBER = {4},
    ARTICLE-NUMBER = {616},
    ISSN = {2072-4292},
    url = {https://doi.org/10.3390/rs10040616}
}

@article{martins2002errors,
    author = {Martins, José Vanderlei and Tanré, Didier and Remer, Lorraine and Kaufman, Yoram and Mattoo, Shana and Levy, Robert},
    title = {{MODIS Cloud screening for remote sensing of aerosols over oceans using spatial variability}},
    journal = {Geophysical Research Letters},
    volume = {29},
    number = {12},
    pages = {MOD4-1-MOD4-4},
    url = {https://doi.org/10.1029/2001GL013252},
    year = {2002}
}

@incollection{heidinger2020cvsbook,
    title = {{7.07 - Satellite Remote Sensing of Cloud Vertical Structure}},
    editor = {Shunlin Liang},
    booktitle = {Comprehensive Remote Sensing (Second Edition)},
    publisher = {Elsevier},
    edition = {Second Edition},
    address = {Oxford},
    pages = {209-247},
    year = {2026},
    isbn = {978-0-443-23949-6},
    url = {https://doi.org/10.1016/B978-0-443-13220-9.00054-8},
    author={Heidinger, AK and Li, Y and Wanzong, S and Noh, Y-J and Walther, A and Tushaus, S and Miller, S},
}

@article{maddux2010geometry,
    author={Maddux, BC and Ackerman, SA and Platnick, S},
    title = {{Viewing Geometry Dependencies in MODIS Cloud Products}},
    journal = "Journal of Atmospheric and Oceanic Technology",
    year = "2010",
    publisher = "American Meteorological Society",
    address = "Boston MA, USA",
    volume = "27",
    number = "9",
    url = {https://doi.org/10.1175/2010JTECHA1432.1},
    pages="1519 - 1528",
}

@article{winker2010calipso,
   author={Winker, DM and Pelon, Jacques and Coakley Jr, JA and Ackerman, SA and Charlson, RJ and Colarco, PR and Flamant, P and Fu, Qinjun and Hoff, RM and Kittaka, C and others},
    title = {{The CALIPSO Mission: A Global 3D View of Aerosols and Clouds}},
    journal = "Bulletin of the American Meteorological Society",
    year = "2010",
    publisher = "American Meteorological Society",
    address = "Boston MA, USA",
    volume = "91",
    number = "9",
    url = {https://doi.org/10.1175/2010BAMS3009.1},
    pages="1211 - 1230",
}

@article{stephens2002cloudsat,
    author={Stephens, Graeme L and Vane, Deborah G and Boain, Ronald J and Mace, Gerald G and Sassen, Kenneth and Wang, Zhien and Illingworth, Anthony J and O'connor, Ewan J and Rossow, William B and Durden, Stephen L and others},
    title = {{THE CLOUDSAT MISSION AND THE A-TRAIN: A New Dimension of Space-Based Observations of Clouds and Precipitation}},
    journal = "Bulletin of the American Meteorological Society",
    year = "2002",
    publisher = "American Meteorological Society",
    address = "Boston MA, USA",
    volume = "83",
    number = "12",
    url = {https://doi.org/10.1175/BAMS-83-12-1771}
}

@article{wehr2023ectech,
    AUTHOR = {Wehr, T. and Kubota, T. and Tzeremes, G. and Wallace, K. and Nakatsuka, H. and Ohno, Y. and Koopman, R. and Rusli, S. and Kikuchi, M. and Eisinger, M. and Tanaka, T. and Taga, M. and Deghaye, P. and Tomita, E. and Bernaerts, D.},
    TITLE = {{The EarthCARE mission -- science and system overview}},
    JOURNAL = {Atmospheric Measurement Techniques},
    VOLUME = {16},
    YEAR = {2023},
    NUMBER = {15},
    PAGES = {3581--3608},
    url = {https://doi.org/10.5194/amt-16-3581-2023}
}

@article{illingsworth2015ec,
    author={Illingworth, Anthony J and Barker, HW and Beljaars, A and Ceccaldi, Marie and Chepfer, H and Clerbaux, Nicolas and Cole, J and Delano{\"e}, Julien and Domenech, C and Donovan, David P and others},
    title = {{The EarthCARE Satellite: The Next Step Forward in Global Measurements of Clouds, Aerosols, Precipitation, and Radiation}},
    journal = "Bulletin of the American Meteorological Society",
    year = "2015",
    publisher = "American Meteorological Society",
    address = "Boston MA, USA",
    volume = "96",
    number = "8",
    url = {https://doi.org/10.1175/BAMS-D-12-00227.1},
    pages="1311 - 1332",
}

@article{watts2011oca,
    author = {Watts, P. D. and Bennartz, R. and Fell, F.},
    title = {{Retrieval of two-layer cloud properties from multispectral observations using optimal estimation}},
    journal = {Journal of Geophysical Research: Atmospheres},
    volume = {116},
    number = {D16},
    pages = {},
    url = {https://doi.org/10.1029/2011JD015883},
    year = {2011}
}

@misc{heidinger2020achaatbd,
    title={{Enterprise AWG Cloud Height Algorithm (ACHA) - Version 3.4}},
    author={Heidinger, Andrew K. and Li ,Yue and Wanzong, Steve},
    year={2020},
    publisher={NOAA NESDIS, Center for Satellite Applications and Research},
    url={https://www.star.nesdis.noaa.gov/jpss/documents/ATBD/ATBD_EPS_Cloud_ACHA_v3.4.pdf}
}

@article{haynes2022lowcloud,
    author={Haynes, John M and Noh, Yoo-Jeong and Miller, Steven D and Haynes, Katherine D and Ebert-Uphoff, Imme and Heidinger, Andrew},
    title = {{Low Cloud Detection in Multilayer Scenes Using Satellite Imagery with Machine Learning Methods}},
    journal = "Journal of Atmospheric and Oceanic Technology",
    year = "2022",
    publisher = "American Meteorological Society",
    address = "Boston MA, USA",
    volume = "39",
    number = "3",
    url = {https://doi.org/10.1175/JTECH-D-21-0084.1},
    pages = "319 - 334",
}

@article{wang2023cloud3d,
    author={Wang, Yiding and Gong, Jie and Wu, Dong L. and Ding, Leah},
    journal={IEEE Transactions on Geoscience and Remote Sensing}, 
    title={{Toward Physics-Informed Neural Networks for 3-D Multilayer Cloud Mask Reconstruction}}, 
    year={2023},
    volume={61},
    number={},
    pages={1-14},
    url={https://doi.org/10.1109/TGRS.2023.3329649}
}

@article{bruning2024cloud3d,
    AUTHOR = {Br\"uning, S. and Niebler, S. and Tost, H.},
    TITLE = {{Artificial intelligence (AI)-derived 3D cloud tomography from geostationary 2D satellite data}},
    JOURNAL = {Atmospheric Measurement Techniques},
    VOLUME = {17},
    YEAR = {2024},
    NUMBER = {3},
    PAGES = {961--978},
    url = {https://doi.org/10.5194/amt-17-961-2024}
}

@article{jeggle2025cloud3d,
    author={Jeggle, Kai and Czerkawski, Mikolaj and Serva, Federico and Le Saux, Bertrand and Neubauer, David and Lohmann, Ulrike},
    title = {{IceCloudNet: 3D Reconstruction of Cloud Ice from Meteosat SEVIRI}},
    journal = "Artificial Intelligence for the Earth Systems",
    year = "2025",
    publisher = "American Meteorological Society",
    address = "Boston MA, USA",
    volume = "4",
    number = "4",
    url = {https://doi.org/10.1175/AIES-D-24-0098.1},
    pages="240098",
}

@Article{amell2024cloud3d,
    AUTHOR = {Amell, A. and Pfreundschuh, S. and Eriksson, P.},
    TITLE = {{The Chalmers Cloud Ice Climatology: retrieval implementation and validation}},
    JOURNAL = {Atmospheric Measurement Techniques},
    VOLUME = {17},
    YEAR = {2024},
    NUMBER = {14},
    PAGES = {4337--4368},
    url = {https://doi.org/10.5194/amt-17-4337-2024},
}

@misc{girtsou2025cloud3d,
    title={{3D Cloud reconstruction through geospatially-aware Masked Autoencoders}}, 
    author={Girtsou, Stella and Salas-Porras, Emiliano Diaz and Freischem, Lilli and Massant, Joppe and Bintsi, Kyriaki-Margarita and Castiglione, Guiseppe and Jones, William and Eisinger, Michael and Johnson, Emmanuel and Jungbluth, Anna},
    year={2025},
    eprint={2501.02035},
    archivePrefix={arXiv},
    primaryClass={cs.CV},
    url ={https://doi.org/10.48550/arXiv.2501.02035}
}

@article{sassen2008cldclasslidar,
    author = {Sassen, Kenneth and Wang, Zhien},
    title = {{Classifying clouds around the globe with the CloudSat radar: 1-year of results}},
    journal = {Geophysical Research Letters},
    volume = {35},
    number = {4},
    pages = {},
    url = {https://doi.org/10.1029/2007GL032591},
    year = {2008}
}

@article{comstock2002cirrus,
    author = {Comstock, Jennifer M. and Ackerman, Thomas P. and Mace, Gerald G.},
    title = {{Ground-based lidar and radar remote sensing of tropical cirrus clouds at Nauru Island: Cloud statistics and radiative impacts}},
    journal = {Journal of Geophysical Research: Atmospheres},
    volume = {107},
    number = {D23},
    pages = {AAC 16-1-AAC 16-14},
    url = {https://doi.org/10.1029/2002JD002203},
    year = {2002}
}

@Article{mason2024comparison,
    AUTHOR = {Mason, S. L. and Barker, H. W. and Cole, J. N. S. and Docter, N. and Donovan, D. P. and Hogan, R. J. and H\"unerbein, A. and Kollias, P. and Puigdom\`enech Treserras, B. and Qu, Z. and Wandinger, U. and van Zadelhoff, G.-J.},
    TITLE = {{An intercomparison of EarthCARE cloud, aerosol, and precipitation retrieval products}},
    JOURNAL = {Atmospheric Measurement Techniques},
    VOLUME = {17},
    YEAR = {2024},
    NUMBER = {2},
    PAGES = {875--898},
    url = {https://doi.org/10.5194/amt-17-875-2024}
}

@article{delanoe2010dardar,
    author = {Delanoë, Julien and Hogan, Robin J.},
    title = {{Combined CloudSat-CALIPSO-MODIS retrievals of the properties of ice clouds}},
    journal = {Journal of Geophysical Research: Atmospheres},
    volume = {115},
    number = {D4},
    pages = {},
    url = {https://doi.org/10.1029/2009JD012346},
    year = {2010}
}

@article{knapp2011gridsat,
    author={Knapp, Kenneth R and Ansari, Steve and Bain, Caroline L and Bourassa, Mark A and Dickinson, Michael J and Funk, Chris and Helms, Chip N and Hennon, Christopher C and Holmes, Christopher D and Huffman, George J and Kossin, James P and Lee, Hai-Tien and Loew, Alexander and Magnusdottir, Gudrun},
    title = "Globally Gridded Satellite Observations for Climate Studies",
    journal = "Bulletin of the American Meteorological Society",
    year = "2011",
    publisher = "American Meteorological Society",
    address = "Boston MA, USA",
    volume = "92",
    number = "7",
    url = {https://doi.org/10.1175/2011BAMS3039.1},
    pages= "893 - 907",
}

@article{schmit2017abi,
      author={Schmit, Timothy J and Griffith, Paul and Gunshor, Mathew M and Daniels, Jaime M and Goodman, Steven J and Lebair, William J},
      title = {{A Closer Look at the ABI on the GOES-R Series}},
      journal = "Bulletin of the American Meteorological Society",
      year = "2017",
      publisher = "American Meteorological Society",
      address = "Boston MA, USA",
      volume = "98",
      number = "4",
      url = {https://doi.org/10.1175/BAMS-D-15-00230.1},
      pages="681 - 698",
}

@article{schmetz2002seviri,
    author={Schmetz, Johannes and Pili, Paolo and Tjemkes, Stephen and Just, Dieter and Kerkmann, Jochen and Rota, Sergio and Ratier, Alain},
    title = {{AN INTRODUCTION TO METEOSAT SECOND GENERATION (MSG)}},
    journal = "Bulletin of the American Meteorological Society",
    year = "2002",
    publisher = "American Meteorological Society",
    address = "Boston MA, USA",
    volume = "83",
    number = "7",
    url = {https://doi.org/10.1175/1520-0477(2002)083<0977:AITMSG>2.3.CO;2},
    pages= "977 - 992",
}

@article{bessho2016ahi,
    title={{An introduction to Himawari-8/9—Japan’s new-generation geostationary meteorological satellites}},
    author={Bessho, Kotaro and Date, Kenji and Hayashi, Masahiro and Ikeda, Akio and Imai, Takahito and Inoue, Hidekazu and Kumagai, Yukihiro and Miyakawa, Takuya and Murata, Hidehiko and Ohno, Tomoo and others},
    journal={Journal of the Meteorological Society of Japan. Ser. II},
    volume={94},
    number={2},
    pages={151--183},
    year={2016},
    publisher={Meteorological Society of Japan},
    url = {https://doi.org/10.2151/jmsj.2016-009}
}

@misc{ncep2015gfs025,
    author       = {{NCEP}},
    title        = {{National Centers for Environmental Prediction GFS 0.25 Degree Global Forecast Grids Historical Archive}},
    publisher    = {{NSF National Center for Atmospheric Research}},
    year         = {2015},
    date         = {2015-01-26},
    url          = {https://doi.org/10.5065/D65D8PWK},
    note         = {Accessed 10 March 2025}
}

@article{mason2023acmcap,
    AUTHOR = {Mason, S. L. and Hogan, R. J. and Bozzo, A. and Pounder, N. L.},
    TITLE = {{A unified synergistic retrieval of clouds, aerosols, and precipitation from EarthCARE: the ACM-CAP product}},
    JOURNAL = {Atmospheric Measurement Techniques},
    VOLUME = {16},
    YEAR = {2023},
    NUMBER = {13},
    PAGES = {3459--3486},
    url = {https://doi.org//10.5194/amt-16-3459-2023}
}

@Article{mason2017captivate,
    AUTHOR = {Mason, S. L. and Chiu, J. C. and Hogan, R. J. and Tian, L.},
    TITLE = {{Improved rain rate and drop size retrievals from airborne Doppler radar}},
    JOURNAL = {Atmospheric Chemistry and Physics},
    VOLUME = {17},
    YEAR = {2017},
    NUMBER = {18},
    PAGES = {11567--11589},
    url = {https://doi.org/10.5194/acp-17-11567-2017}
}

@article{mason2018captivate,
    author = {Mason, S. L. and Chiu, C. J. and Hogan, R. J. and Moisseev, D. and Kneifel, S.},
    title = {Retrievals of Riming and Snow Density From Vertically Pointing Doppler Radars},
    journal = {Journal of Geophysical Research: Atmospheres},
    volume = {123},
    number = {24},
    pages = {13,807-13,834},
    url = {https://doi.org/10.1029/2018JD028603},
    year = {2018}
}

@misc{heidinger2020ecm,
    title={{NOAA Enterpreise Cloud Mask - Version 1.2}},
    author={Heidinger, Andrew K. and Straka, William},
    year={2020},
    publisher={NOAA NESDIS, Center for Satellite Applications and Research},
    url={https://www.star.nesdis.noaa.gov/goesr/documents/ATBDs/Enterprise/ATBD_Enterprise_Cloud_Mask_v1.2_2020_10_01.pdf}
}

@misc{pavolonis2020phase,
    title={{Enterprise Algorithm Theoretical Basis Document For Cloud Type and Cloud Phase}},
    author={Pavolonis, Michael and Calvert, Corey},
    year={2020},
    publisher={NOAA NESDIS, Center for Satellite Applications and Research},
    url={https://www.star.nesdis.noaa.gov/JPSS/documents/ATBD/ATBD_EPS_Cloud_CldType_v3.0.pdf}
}

@misc{wang2011microbase,
    title={{Continuous Baseline Microphysical Retrieval (MICROBASEKAPLUS), 2024-01-01 to 2024-06-30, Southern Great Plains (SGP), Central Facility, Lamont, OK (C1)}},
    year="2011",
    url = {https://doi.org/10.5439/1768890},
    journal={Atmospheric Radiation Measurement (ARM) user facility},
    author={Wang, Meng and Giangrande, Scott and Johnson, Karen and Jensen, Michael}
}

@mastersthesis{smalley2011srf,
    author       = {Smalley, Mark Allen},
    title        = {{Effects of Spectral Response Function Differences on {CO2} Slicing with an Application to Cloud Climatologies}},
    school       = {University of Wisconsin--Madison},
    year         = {2011},
    type         = {Master's thesis},
    department   = {Department of Atmospheric and Oceanic Sciences},
    journal      = {Journal of the UW-AOS},
    volume       = {16},
    url = {https://www.aos.wisc.edu/uwaosjournal/Volume16/MarkSmalley_MS_Thesis.pdf}
}

@article{baum2012srf,
    author={Baum, Bryan A and Menzel, W Paul and Frey, Richard A and Tobin, David C and Holz, Robert E and Ackerman, Steve A and Heidinger, Andrew K and Yang, Ping},
    title = {{MODIS Cloud-Top Property Refinements for Collection 6}},
    journal = "Journal of Applied Meteorology and Climatology",
    year = "2012",
    publisher = "American Meteorological Society",
    address = "Boston MA, USA",
    volume = "51",
    number = "6",
    url = {https://doi.org/10.1175/JAMC-D-11-0203.1},
    pages="1145 - 1163",
}

@article{meyer2020srf,
    AUTHOR = {Meyer, Kerry and Platnick, Steven and Holz, Robert and Dutcher, Steve and Quinn, Greg and Nagle, Fred},
    TITLE = {{Derivation of Shortwave Radiometric Adjustments for SNPP and NOAA-20 VIIRS for the NASA MODIS-VIIRS Continuity Cloud Products}},
    JOURNAL = {Remote Sensing},
    VOLUME = {12},
    YEAR = {2020},
    NUMBER = {24},
    ARTICLE-NUMBER = {4096},
    ISSN = {2072-4292},
    url = {https://doi.org/10.3390/rs12244096}
}

@misc{perez2017film,
    title={{FiLM: Visual Reasoning with a General Conditioning Layer}}, 
    author={Perez, Ethan and Strub, Florian and De Vries, Harm and Dumoulin, Vincent and Courville, Aaron},
    year={2017},
    eprint={1709.07871},
    archivePrefix={arXiv},
    primaryClass={cs.CV},
    url={https://doi.org/10.48550/arXiv.1709.07871}
}

@misc{liu2022convnext,
    title={{A ConvNet for the 2020s}}, 
    author={Liu, Zhuang and Mao, Hanzi and Wu, Chao-Yuan and Feichtenhofer, Christoph and Darrell, Trevor and Xie, Saining},
    year={2022},
    eprint={2201.03545},
    archivePrefix={arXiv},
    primaryClass={cs.CV},
    url={https://doi.org/10.48550/arXiv.2201.03545}
}

@INPROCEEDINGS{vandal2025earthnet,
    author={Vandal, Thomas J. and Duffy, Kate and Nachmany, Yoni and McDuff, Daniel},
    booktitle={{2025 IEEE International Conference on Data Mining Workshops (ICDMW)}}, 
    title={{Global Atmospheric Data Assimilation with Multi-Modal Masked Autoencoders}}, 
    year={2025},
    volume={},
    number={},
    pages={863-872},
    url={https://doi.org/10.1109/ICDMW69685.2025.00103}
}

@misc{loschilov2019adamw,
    title={{Decoupled Weight Decay Regularization}}, 
    author={Loshchilov, Ilya and Hutter, Frank},
    year={2019},
    eprint={1711.05101},
    archivePrefix={arXiv},
    primaryClass={cs.LG},
    url={https://arxiv.org/abs/1711.05101}, 
}

@misc{kingma2017adam,
    title={{Adam: A Method for Stochastic Optimization}}, 
    author={Kingma, Diederik P and Ba, Jimmy},
    year={2017},
    eprint={1412.6980},
    archivePrefix={arXiv},
    primaryClass={cs.LG},
    url={https://arxiv.org/abs/1412.6980}, 
}

@article{schulte2023cloudsat,
    AUTHOR = {Schulte, R. M. and Lebsock, M. D. and Haynes, J. M.},
    TITLE = {{What CloudSat cannot see: liquid water content profiles inferred from MODIS and CALIOP observations}},
    JOURNAL = {Atmospheric Measurement Techniques},
    VOLUME = {16},
    YEAR = {2023},
    NUMBER = {14},
    PAGES = {3531--3546},
    url = {https://doi.org/10.5194/amt-16-3531-2023}
}

@ARTICLE{dimichele2013cthlidar,
    author={Di Michele, Sabatino and McNally, Tony and Bauer, Peter and Genkova, Iliana},
    journal={IEEE Transactions on Geoscience and Remote Sensing}, 
    title={{Quality Assessment of Cloud-Top Height Estimates From Satellite IR Radiances Using the CALIPSO Lidar}}, 
    year={2013},
    volume={51},
    number={4},
    pages={2454-2464},
    url={https://doi.org/10.1109/TGRS.2012.2210721}
}

@misc{walther2013atbd,
    title={{ABI Algorithm Theoretical Basis Document For Daytime Cloud Optical and Microphysical Properties (DCOMP)}},
    author={Walther, Andi and Straka, William and Heidinger, Andrew K.},
    year={2013},
    publisher={NOAA NESDIS, Center for Satellite Applications and Research},
    url ={https://www.star.nesdis.noaa.gov/goesr/documents/ATBDs/Baseline/ATBD_GOES-R_Cloud_DCOMP_v3.0_Jun2013.pdf}
}

@misc{jeggle2023spatial,
    title={{IceCloudNet: Cirrus and mixed-phase cloud prediction from SEVIRI input learned from sparse supervision}}, 
    author={Jeggle, Kai and Czerkawski, Mikolaj and Serva, Federico and Saux, Bertrand Le and Neubauer, David and Lohmann, Ulrike},
    year={2023},
    eprint={2310.03499},
    archivePrefix={arXiv},
    primaryClass={physics.ao-ph},
    url={https://arxiv.org/abs/2310.03499}, 
}

@article{hilburn2023spatial,
    author = {Hilburn, Kyle A.},
    title = {{Understanding Spatial Context in Convolutional Neural Networks Using Explainable Methods: Application to Interpretable GREMLIN}},
    journal = "Artificial Intelligence for the Earth Systems",
    year = "2023",
    publisher = "American Meteorological Society",
    address = "Boston MA, USA",
    volume = "2",
    number = "3",
    url = {https://doi.org/10.1175/AIES-D-22-0093.1},
    pages= "220093",
}

@article{white2025unetcomp,
    author = {White, Charles H. and Noh, Yoo-Jeong and Haynes, John M. and Ebert-Uphoff, Imme},
    title = {{Emulating Daytime ABI Cloud Optical Properties at Night With Machine Learning}},
    journal = {Journal of Geophysical Research: Atmospheres},
    volume = {130},
    number = {12},
    pages = {e2024JD042829},
    url = {https://doi.org/10.1029/2024JD042829},
    year = {2025}
}

@Article{noh2022cbhaviation,
    AUTHOR = {Noh, Yoo-Jeong and Haynes, John M. and Miller, Steven D. and Seaman, Curtis J. and Heidinger, Andrew K. and Weinrich, Jeffrey and Kulie, Mark S. and Niznik, Mattie and Daub, Brandon J.},
    TITLE = {{A Framework for Satellite-Based 3D Cloud Data: An Overview of the VIIRS Cloud Base Height Retrieval and User Engagement for Aviation Applications}},
    JOURNAL = {Remote Sensing},
    VOLUME = {14},
    YEAR = {2022},
    NUMBER = {21},
    ARTICLE-NUMBER = {5524},
    ISSN = {2072-4292},
    url = {https://doi.org/10.3390/rs14215524}
}

@INPROCEEDINGS{noh2024cloud3d,
    author={Noh, Y. J. and Haynes, J. M. and Miller, S. D. and White, C. and Daub, B. and Ebert-Uphoff, I. and Yu, H. and Apke, J. and Chase, R. and Haynes, K. and Solbrig, J. and Rose, E. and King, M. and Cheatwood, L.},
    booktitle={IGARSS 2024 - 2024 IEEE International Geoscience and Remote Sensing Symposium}, 
    title={{Building Unified Global 3D Cloud Data From Multiple Satellites For Advancing Weather And Climate Research}}, 
    year={2024},
    volume={},
    number={},
    pages={1280-1283},
    url={https://doi.org/10.1109/IGARSS53475.2024.10642296}
  }

@article{bernstein2005icing,
    author={Bernstein, Ben C and McDonough, Frank and Politovich, Marcia K and Brown, Barbara G and Ratvasky, Thomas P and Miller, Dean R and Wolff, Cory A and Cunning, Gary},
    title = {{Current Icing Potential: Algorithm Description and Comparison with Aircraft Observations}},
    journal = "Journal of Applied Meteorology",
    year = "2005",
    publisher = "American Meteorological Society",
    address = "Boston MA, USA",
    volume = "44",
    number = "7",
    url = {https://doi.org/10.1175/JAM2246.1},
    pages= "969 - 986",
}

@techreport{jeck2001icing,
    author      = {Jeck, Richard K.},
    title       = {A History and Interpretation of Aircraft Icing Intensity Definitions and {FAA} Rules for Operating in Icing Conditions},
    institution = {Federal Aviation Administration William J. Hughes Technical Center},
    year        = {2001},
    number      = {DOT/FAA/AR-01/91},
    address     = {Atlantic City International Airport, NJ},
    url         = {https://www.faa.gov/sites/faa.gov/files/aircraft/air_cert/design_approvals/small_airplanes/aceReportAR-01-91.pdf}
}

@Article{eisinger2024earthcare,
    AUTHOR = {Eisinger, M. and Marnas, F. and Wallace, K. and Kubota, T. and Tomiyama, N. and Ohno, Y. and Tanaka, T. and Tomita, E. and Wehr, T. and Bernaerts, D.},
    TITLE = {{The EarthCARE mission: science data processing chain overview}},
    JOURNAL = {Atmospheric Measurement Techniques},
    VOLUME = {17},
    YEAR = {2024},
    NUMBER = {2},
    PAGES = {839--862},
    url = {https://doi.org/10.5194/amt-17-839-2024}
}

@article{kubota2026earthcare,
    title = {{Summary of the EarthCARE science and validation workshop 2025 -Early achievements and future perspectives for the Earth Cloud Aerosol and Radiation Explorer (EarthCARE) satellite mission-}},
    journal = {Journal of the European Meteorological Society},
    volume = {4},
    pages = {100037},
    year = {2026},
    issn = {2950-6301},
    url = {https://doi.org/10.1016/j.jemets.2026.100037},
    author={Kubota, Takuji and Kikuchi, Maki and Muto, Masataka and Hashimoto, Makiko and Imura, Yuki and Maruyama, Kenta and Hoffmann, Alex and Hummel, Timon and Koopman, Robert and Malina, Edward and others},
}

@article{han1994cwp,
    author={Han, Qingyuan and Rossow, William B and Lacis, Andrew A},
    title = {{Near-Global Survey of Effective Droplet Radii in Liquid Water Clouds Using ISCCP Data}},
    journal = "Journal of Climate",
    year = "1994",
    publisher = "American Meteorological Society",
    address = "Boston MA, USA",
    volume = "7",
    number = "4",
    url = {https://doi.org/10.1175/1520-0442(1994)007<0465:NGSOED>2.0.CO;2},
    pages="465 - 497",
}

@article{bennartz2007cwp,
    author = {Bennartz, R.},
    title = {{Global assessment of marine boundary layer cloud droplet number concentration from satellite}},
    journal = {Journal of Geophysical Research: Atmospheres},
    volume = {112},
    number = {D2},
    pages = {},
    url = {https://doi.org/10.1029/2006JD007547},
    year = {2007}
}

@article{wood2006spatial,
    author={Wood, Robert and Hartmann, Dennis L},
    title = {{Spatial Variability of Liquid Water Path in Marine Low Cloud: The Importance of Mesoscale Cellular Convection}},
    journal = "Journal of Climate",
    year = "2006",
    publisher = "American Meteorological Society",
    address = "Boston MA, USA",
    volume = "19",
    number = "9",
    url = {https://doi.org/10.1175/JCLI3702.1},
    pages= "1748 - 1764"
}

\end{document}